%% file: main.tex
\documentclass[preprint,12pt]{elsarticle}

\usepackage[utf8]{inputenc}
\usepackage[T1]{fontenc}

\usepackage{times} 
\usepackage{microtype} 

\usepackage{graphicx}

\usepackage{booktabs}
\usepackage{longtable}
\usepackage{tabularx}
\usepackage{multirow}
\usepackage{makecell}
\usepackage{array}

\usepackage{amsmath}
\usepackage{amsthm}
\usepackage{algorithm}
\usepackage{algorithmic}

\usepackage{soul}
\usepackage{url}
\usepackage{inconsolata} 
\usepackage{rotating}
\usepackage{pdflscape}
\usepackage[switch]{lineno}
\usepackage{pifont}
\usepackage{enumitem}

\usepackage{xcolor}

\journal{Journal of Biomedical Informatics}

\begin{document}
\begin{frontmatter}
\title{Before and After ChatGPT: Revisiting AI-Based Dialogue Systems for Emotional Support}

\author[1,2]{Daeun Lee\fnref{fn1}}
\ead{da-eun.lee@yale.edu}

\author[3]{Dongje Yoo\fnref{fn1}}
\ead{foryui25@skku.edu}

\author[1]{Migyeong Yang}
\ead{mgyang@g.skku.edu}

\author[4]{Jihyun An}
\ead{jh85.an@samsung.com}

\author[2]{Christine B. Cha}
\ead{christine.cha@yale.edu}

\author[1]{Jinyoung Han\corref{cor1}}
\ead{jinyounghan@skku.edu}
\cortext[cor1]{Corresponding author.}

\fntext[fn1]{These authors contributed equally to this work.}

\address[1]{Department of Applied Artificial Intelligence, Sungkyunkwan University, Seoul, Republic of Korea}
\address[2]{Yale School of Medicine, Yale University, New Haven, CT, USA}
\address[3]{Department of Immersive Media Engineering, Sungkyunkwan University, Seoul, Republic of Korea}
\address[4]{Department of Psychiatry, Samsung Medical Center, Seoul, Republic of Korea}

\begin{abstract}
\input{0-abstract}

\end{abstract}

\begin{highlights}
\item An analysis of 146 studies indicates a steady rise in research on AI-based mental health dialogue systems since 2020.
\item Prior to the introduction of ChatGPT, research largely relied on specialized neural models optimized for specific tasks and equipped with explicit empathy and knowledge mechanisms.
\item In contrast, post-ChatGPT studies increasingly adopt general-purpose conversational models, offering improved linguistic flexibility but revealing ongoing concerns related to safety and reliability.
\end{highlights}


\end{frontmatter}

\input{1-introduction}
\input{3-method}

\input{4-experiment1-biblio}
\input{4-experiment2-top10}

\input{4-experiment3-ESconv}
\input{4-experiment4-LLM}
\input{5-Disucssion}

\input{6-Conclusion}

\input{7-ack}

\bibliographystyle{elsarticle-num-names} 
\bibliography{reference}

\end{document}

%% file: 0-abstract.tex
\noindent\textbf{Background}:
Mental health remains a major public health concern, while access to timely psychological support is often limited. AI-based dialogue systems have emerged as promising tools to address these barriers, and recent advances in large language models (LLMs) have substantially transformed this research area. However, a systematic understanding of this technological transition is still lacking.

\noindent\textbf{Objective}:
This study aims to review the technological transition in AI-driven dialogue systems for mental health, focusing on the shift from task-specific deep learning models to LLM-based approaches.

\noindent\textbf{Methods}:
We conducted a bibliometric analysis and a qualitative trend review of studies published between 2020 and May 2024 using Web of Science, Scopus, and the ACM Digital Library. The qualitative analysis examined the technological transition by comparing studies conducted before and after the widespread adoption of LLMs (e.g., ChatGPT), with pre-LLM research represented by highly cited studies and work based on the ESConv dataset, and post-LLM research represented by highly cited LLM-based dialogue systems.

\noindent\textbf{Results}:
A total of 146 studies met the inclusion criteria, revealing a consistent growth in publications over time. Prior to the broad use of LLMs, empathetic response generation largely depended on task-specific deep learning models, particularly in highly cited and ESConv-based studies that focused on multi-task learning and external knowledge integration. In contrast, more recent, highly cited studies adopting LLM-based dialogue systems demonstrated improved linguistic flexibility and generalization for emotional support, while also revealing challenges related to reliability and safety in mental health applications.

\noindent\textbf{Conclusions}:
This review highlights how AI-based dialogue systems for mental health have advanced through the use of LLMs. By clarifying current research trends and limitations, our findings provide guidance for developing more effective and reliable AI-driven counseling systems.

%% file: 1-introduction.tex
\section{Introduction}

Mental disorders have become a critical global public health issue. The OECD (Organization for Economic Cooperation and Development) reports that, in the United States, 14.1 out of every 100,000 people die by suicide each year~\cite{oecd2020}. Unfortunately, over 80\% of these suicides are committed by individuals suffering from mental illness~\cite{stack2014mental}.
Although early intervention is crucial, traditional mental health services, such as counseling and psychological therapy, often encounter significant barriers, including financial constraints, time limitations, and geographic restrictions~\cite{42american, kivlighan2021role}. Furthermore, stigma and shame associated with discussing personal difficulties contribute to the reluctance of many individuals to seek support~\cite{prior2012overcoming}. These challenges became more evident with the increasing demand for digital mental health solutions during the COVID-19 pandemic~\cite{balcombe2021digital}. Advancements in artificial intelligence (AI) have opened new possibilities for addressing these barriers by enhancing the accessibility and scalability of psychological support~\cite{balcombe2021digital,rangaswamy2024ai,trappey2022development}. In particular, progress in natural language processing (NLP) has enabled the development of AI-based conversational agents capable of generating coherent and contextually appropriate interactions with users~\cite{chen2017survey}. These technological developments have positioned dialogue systems as promising tools for mental health interventions.

Early research in this area predominantly relied on task-specific deep learning models designed for narrowly defined objectives, such as emotion recognition or empathetic response generation. More recently, the emergence of large language models (LLMs), trained on extensive corpora, has marked a significant technological transition. LLMs demonstrate an enhanced capacity to generate fluent, human-like responses~\cite{huo2025large} and can function as general-purpose systems adaptable to a wide range of emotional support tasks. Consequently, AI based dialogue systems--often referred to as chatbots--have been increasingly applied in healthcare and mental health contexts~\cite{cho-etal-2023-integrative,shi2024medical}. Despite these advances, substantial limitations remain. LLMs are still prone to \textit{hallucinations}, producing incorrect or irrelevant information in response to user inputs~\cite{ji2023towards,verspoor2024fighting}. In addition, their opaque training data and lack of domain-specific grounding, particularly in mental health contexts, raise concerns about reliability and safety~\cite{agrawal2022large}. As a result, many real-world applications continue to employ rule-based or traditional deep learning systems due to their predictability and controllability~\cite{denecke2021artificial}, albeit at the cost of reduced conversational flexibility and expressiveness~\cite{li2023systematic}. These observations highlight a critical gap between rapid technological advancements and the practical requirements of mental health counseling.

Against this backdrop, this review focuses on the technological transition in AI-based dialogue systems for emotional support, particularly the shift from task-specific deep learning approaches to LLM-based systems. To capture this transition, we concentrate on studies published between 2020 and May 2024, a period that encompasses the introduction of GPT-3 (2020) ~\cite{brown2020language}, the emergence of GPT-3.5 and ChatGPT (2022) ~\cite{roumeliotis2023chatgpt}, and the release of GPT-4 (2023) ~\cite{achiam2023gpt}, which collectively marked a turning point in the application of LLMs to dialogue-based mental health interventions. Studies published prior to 2020 were excluded, as LLMs had not yet been widely adopted for real-world dialogue system applications, and most approaches relied on traditional, task-specific models. The review period was limited to May 2024 to maintain a focus on text-based dialogue systems, as subsequent models introduced real-time multimodal capabilities that extend beyond the scope of this study (i.e., GPT-4o ~\cite{hurst2024gpt}). Within this temporal framework, we qualitatively analyze two major strands of research: (1) task-specific deep learning models explicitly designed for emotional support tasks, and (2) LLM-based approaches adapted to this domain. By examining how research emphases evolved across these periods, we aim to clarify how rapid technological advancements have reshaped dialogue system design and to assess their alignment with the practical and psychological needs of mental health counseling. The Discussion section further explores emerging directions beyond this transition and outlines future research opportunities.

\section*{Statement of Significance:}

%% file: 3-method.tex
\section{Method}

\subsection{Data Sources and Search Strategy}
\begin{figure}[ht]
    \centering
    \includegraphics[width=0.7\linewidth]{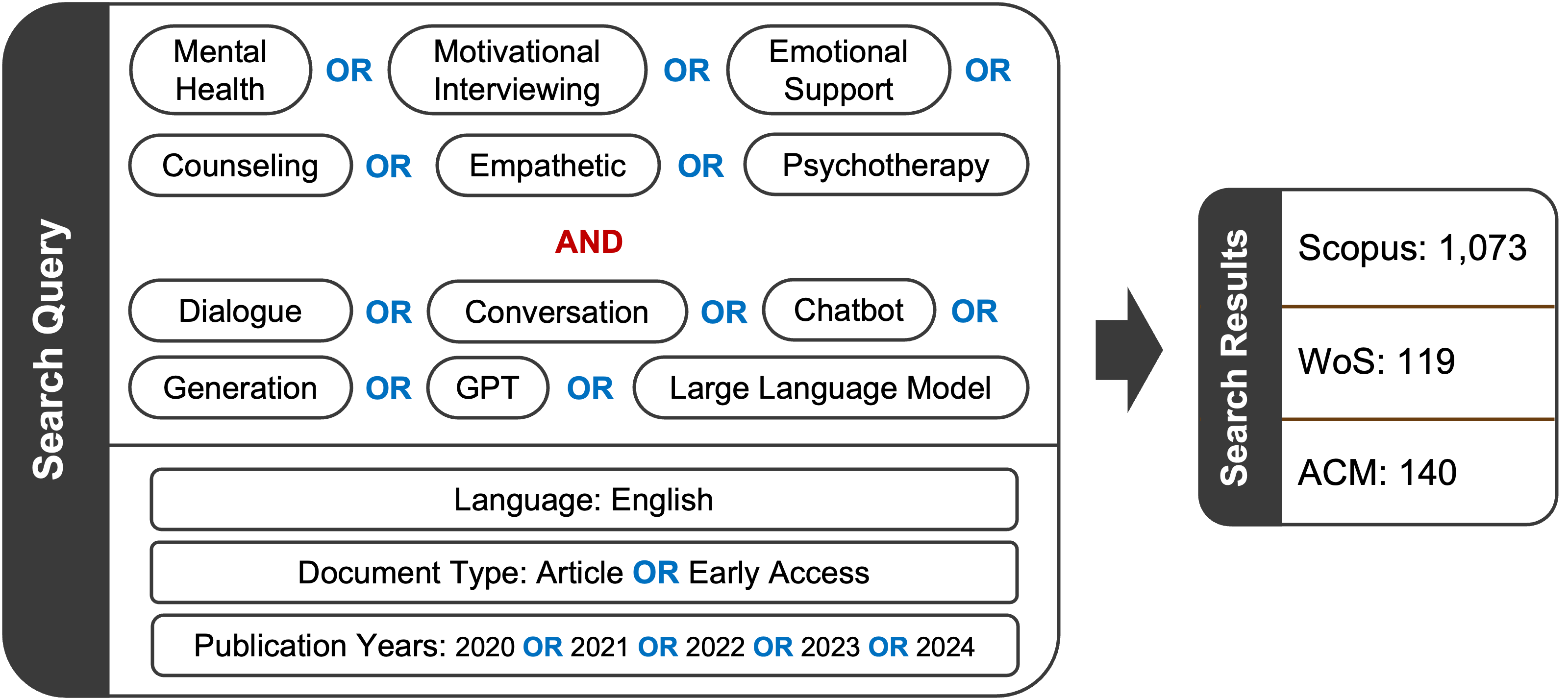} 
    \caption{Search Query Categories with Results}
    \label{fig:searchquery}
\end{figure}

We sourced papers from three major citation databases: Scopus, Web of Science (WoS), and the Association for Computing Machinery (ACM). Scopus~\cite{scopus2024} is one of the largest citation repositories, encompassing a wide range of scientific journals, conference papers, and books. WoS~\cite{wos2024} includes reputable publications categorized under the Science Citation Index Expanded (SCIE), the Social Sciences Citation Index (SSCI), and the Arts \& Humanities Citation Index (A\&HCI). The ACM Digital Library~\cite{acmdl2024}, a prominent organization in computing, provides an extensive digital library that includes journals, conference proceedings, and technical magazines, making it a key resource for research in computer science and information technology. We retrieved relevant publications where the search terms appeared in the title, abstract, or keywords. The search queries were developed based on previous research on conversational agents for mental health~\cite{liu2021towards,rashkin2019towards,wu2022anno}, as illustrated in Figure~\ref{fig:searchquery}. Although PubMed and Google Scholar were considered, they were excluded due to scope and methodological considerations~\cite{kim2021machine}. PubMed predominantly indexes biomedical research with limited representation of computer science conference literature, and Google Scholar does not offer sufficient transparency or reproducibility in search and indexing processes.

The selection criteria for studies included those published in English, appearing in peer-reviewed scientific journals or conference proceedings, which are recognized as high-quality publication venues in engineering and computer science. Studies categorized as closed access were excluded. Papers published between 2020 and 2024 were retrieved as of May 2024.

\subsection{Eligibility Criteria}
Studies were included if they (1) aimed to address technical limitations of existing AI models; (2) focused on developing/applying AI models that automatically generate language-based outputs (e.g., dialogue systems, Question Answering); (3) trained/finetuned AI models; (4) evaluated one or more capabilities for counseling or mental health-related services (e.g., empathy, conversation); and (5) provided a clear explanation of algorithms or code to support reproducibility. We specifically excluded studies that (1) evaluated clinical effectiveness or user experience, or (2) aimed at generating training datasets (e.g., virtual episode generation).

\subsection{Study Screening and Selection}
The search results were imported into a Google Spreadsheet for management. The authors individually screened the titles and abstracts of the studies to assess their eligibility. Studies that met all inclusion criteria proceeded to full-text review for a final selection decision. The authors discussed each screening stage to reach agreement. As a result, out of a total of 1,332 papers retrieved through the database search, 146 papers were ultimately included, as described in Figure~\ref{fig:screening}. The complete list of these 146 papers is available in Supplementary material.

\begin{figure}[tbp]
    \centering
    \includegraphics[width=0.7\linewidth]{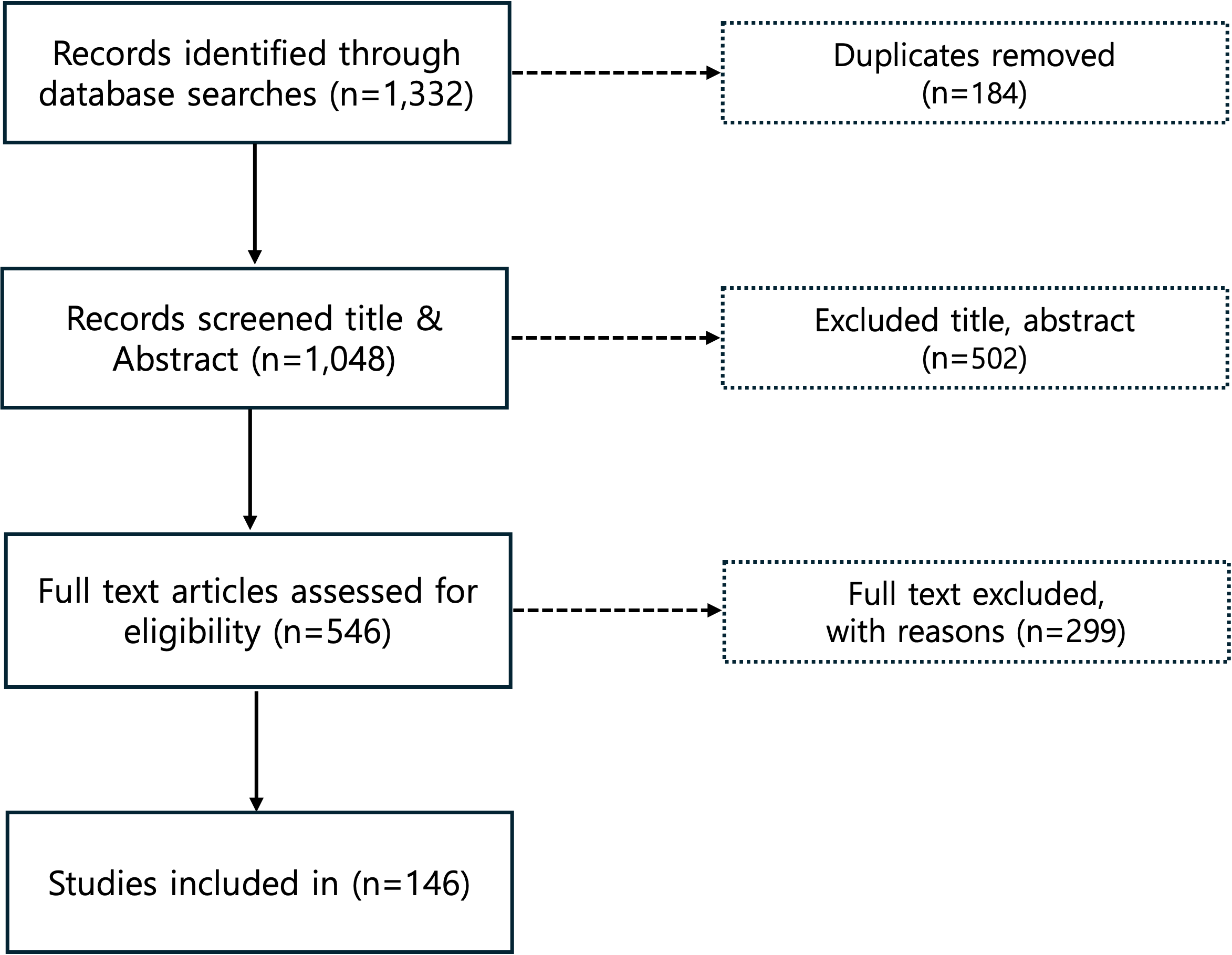} 
    \caption{PRISMA (Preferred Reporting Items for Systematic Reviews and Meta-Analyses) flowchart~\cite{moher2010preferred}}
    \label{fig:screening}
\end{figure}

\subsection{Analysis Methodology}
In this study, we applied bibliometric analysis, a widely employed quantitative method for examining literature, which plays a significant role in AI/ML healthcare research by highlighting developmental trends and ensuring the generation of measurable, consistent, and unbiased results~\cite{kim2021machine}. We first began by analyzing the distribution of publications across different categories, such as sources, countries, institutions, and authors. We also performed a network analysis of commonly used keywords to uncover dominant themes and emerging trends within the literature. Note that statistical evaluations were carried out using Python and Microsoft Excel. Moreover, we conducted a trend analysis across three categories: (i) highly cited publications, (ii) publications utilizing the well-known counseling open dataset, ESConv~\cite{liu2021towards}, and (iii) publications employing LLMs. We examined the characteristics of the papers in each category and discussed the challenges in developing fully functional psychotherapy dialogue systems. While the bibliometric analysis is largely software-generated, the trend analysis is subjective and guided by the authors.

%% file: 4-experiment1-biblio.tex
\section{Result}~\label{sec:biblio}

\subsection{Overall Publication Trend}
Table~\ref{tab:year_table} demonstrates a steady increase in publications from 2020 to 2024 (up to May 2024). In 2020, only 9 papers were screened, whereas publication activity surged in 2023, with 62 papers screened. 

\begin{table}[!ht]
\centering
\resizebox{0.22\textwidth}{!}{%
\begin{tabular}{c|c}
\hline
\textbf{Year} & \textbf{Count, n (\%)} \\
\hline 
2020 & 9 (6.1\%)   \\
2021 & 25 (17.1\%) \\
2022 & 36 (24.7\%) \\
2023 & 62 (42.4\%)  \\
2024 & 14 (9.6\%)  \\ \hline
\textbf{Total} & \textbf{146 (100.0\%)}  \\
\hline
\end{tabular}
}
\caption{Number of Publications by Year}
\label{tab:year_table}
\end{table}

\noindent\textbf{\textit{Productive Publication Source}}

We examined the publication sources of the selected papers, which included journal articles, conference proceedings, and book chapters. Table~\ref{tab:publication_table} displays the sources with the highest number of publications across Scopus, WoS, and ACM. Proceedings of the Annual Meeting of the Association for Computational Linguistics (ACL), a top-tier natural language processing conference in computer science, was identified as the leading source, closely followed by Lecture Notes in Computer Science, with both contributing more than 10 publications.

\begin{table}[!ht]
\centering
\resizebox{\textwidth}{!}{%
\begin{tabular}{c|c|c}
\hline
\textbf{Rank} & \textbf{Source}                                                                     & \textbf{Count, n (\%)} \\ \hline
1    & Annual Meeting of the Association for Computational Linguistics (ACL)      & 16 (10.9)                         \\
2    & Lecture Notes in Computer Science                                          & 13 (8.9)                          \\
3 & IEEE International Conference on Acoustics, Speech and Signal Processing (ICASSP)                 & 6 (4.1) \\
3    & Findings of the Association for Computational Linguistics (EMNLP Findings) & 6 (4.1)                           \\
5    & Conference on Empirical Methods in Natural Language Processing (EMNLP)     & 5 (3.4)                           \\
6 & Conference of the North American Chapter of the Association for Computational Linguistics (NAACL) & 4 (2.7) \\
7    & AAAI Conference on Artificial Intelligence (AAAI)                          & 4 (2.7)                           \\
8    & Sun SITE Central Europe Workshop (CEUR Workshop)                           & 3 (2.1)                           \\
8    & Knowledge-Based Systems                                                    & 3 (2.1)                           \\
8    & The Web (formerly, World Wide Web Conference)                              & 3 (2.1)                           \\ \hline
\end{tabular}%
}
\caption{Top Sources for Publications}
\label{tab:publication_table}
\end{table}

\noindent\textbf{\textit{Predominant Countries}}

As shown in Table~\ref{tab:country_table}, more than 10 countries were recognized as the most productive based on their publication output. China was the leading contributor, followed by India and the United States.

\begin{table}[!ht]
\centering
\resizebox{0.4\textwidth}{!}{%
\begin{tabular}{c|c|c} 
\hline
\textbf{Rank} & \textbf{Country} & \textbf{Count, n (\%)}\\
\hline
1 & China  & 54 (37.0) \\
2 & India & 25 (17.1) \\
3 & United States & 10 (6.8) \\
3 & Hong Kong & 8 (5.5)\\
5 & South Korea  & 7 (4.8)\\
6 & Japan & 4 (2.7)\\
7 & United Kingdom & 4 (2.7)\\
8 & Taiwan  & 4 (2.7)\\
8 & Australia & 3 (2.1)\\
8 & Singapore & 3 (2.1)\\
\hline
\end{tabular}
}
\caption{Top Countries for Publications}
\label{tab:country_table}
\end{table}

\noindent\textbf{\textit{Productive Institutions}}

A total of 90 institutions were associated with the 146 publications. The top-ranked institutions are listed in Table~\ref{tab:institution_table}. Tsinghua University in China was the most productive institution, followed by the Indian Institute of Technology.

\begin{table}[!ht]
\centering
\resizebox{0.8\textwidth}{!}{%
\begin{tabular}{c|c|c|c} 
\hline
\textbf{Rank} & \textbf{Institution} & \textbf{Country} & \textbf{Count, n (\%)} \\ 
\hline
1 & Tsinghua University & China  & 8 (5.5)\\
2 & Indian Institute of Technology & India  & 7 (4.8)\\
3 & Institute of Information Engineering  & China & 5 (3.4)\\
4 & Shandong University  & China & 4 (2.7)\\
4 & Northeastern University  & United States & 4 (2.7) \\
4 & Harbin Institute of Technology  & China & 4 (2.7)\\
7 & Tianjin University  & China & 3 (2.1) \\
7 & National Cheng Kung University  & Taiwan & 3 (2.1)\\
9 & The University of Tokyo  & Japan & 2 (1.4)\\
9 & Beijing University  & China & 2 (1.4)\\
9 & Seoul National University  & South Korea & 2 (1.4)\\
\hline
\end{tabular}
}
\caption{Top Institutions for Publications}
\label{tab:institution_table}
\end{table}

\noindent\textbf{\textit{Predominant Authors}}

The top 10 researchers in this field are presented in Table~\ref{tab:author_table}, ranked by their publication count. Six of these researchers are affiliated with institutions in China, while two are associated with the Indian Institute of Technology. The most prolific researcher was Professor Su Y from Northwest Normal University, who had three publications.

\begin{table}[!ht]
\centering
\resizebox{0.8\textwidth}{!}{%
\begin{tabular}{c|c|c|c} 
\hline
\textbf{Author} & \textbf{Institution} & \textbf{Country} & \textbf{\textbf{Count, n (\%)}}\\
\hline
Su Y&Northwest Normal University&United States&3 (2.1)\\
Saha T	&Indian Institute of Technology&India&2 (1.4)\\
Bi G&Institute of Information Engineering&China&2 (1.4)\\
Majumder N&Singapore University of Technology and Design&Singapore&2 (1.4)\\
Zhou J&Tsinghua University&China&2 (1.4)\\
Peng W&Institute of Information Engineering&China&2 (1.4)\\
Shen S&University of Michigan&United States&2 (1.4)\\
Mishra K&Indian Institute of Technology&India&2 (1.4)\\
Wang J&Hong Kong Polytechnic University&Hong Kong&2 (1.4)\\
Li Q&Shandong University&China&2 (1.4)\\
\hline
\end{tabular}
}
\caption{Top 10 Most Productive Authors for Publications}
\label{tab:author_table}
\end{table}

\begin{figure}[tbp] 
    \centering 
    \includegraphics[width=0.6\linewidth]{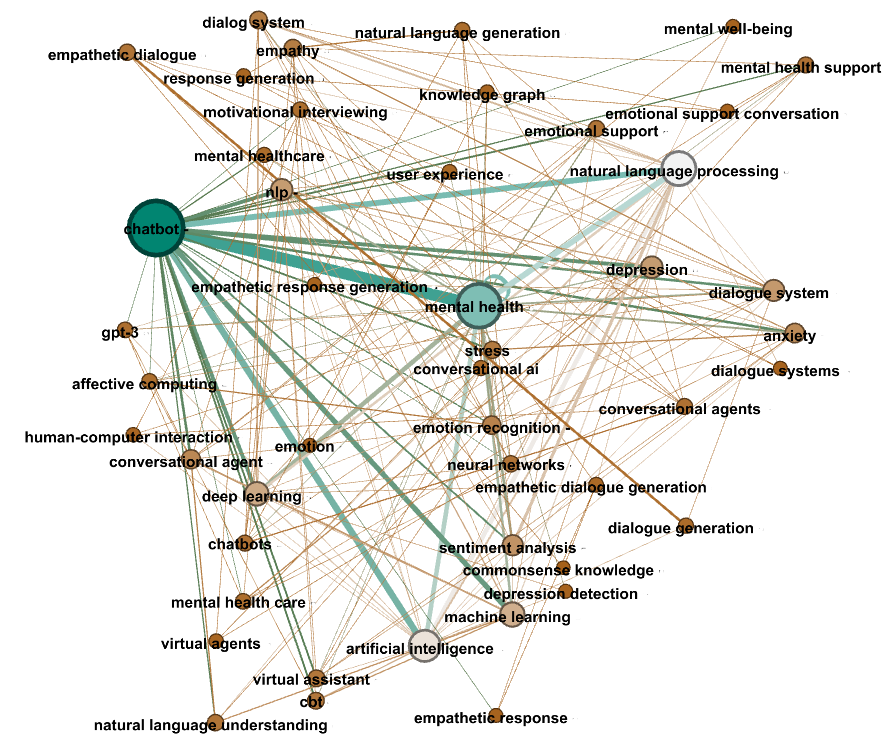}
    \caption{Keyword Co-occurrence Network Graph from 2020 to May 2024}
    \label{fig:network} 
\end{figure}

\begin{figure}[tbp]
    \centering
    \includegraphics[width=0.8\linewidth]{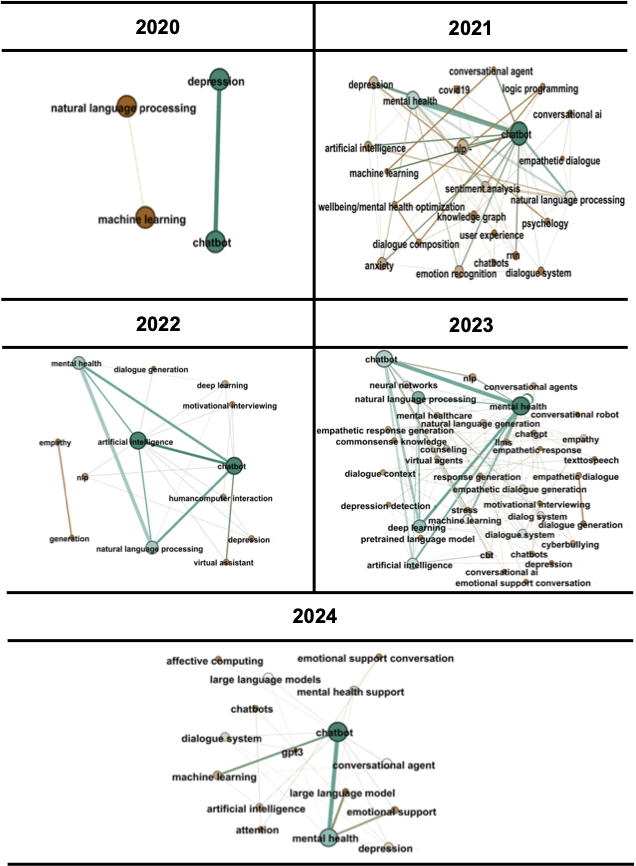} 
    \caption{Yearly Keyword Co-occurrence Network Graphs (2020–2024)}
    \label{fig:year_keyword}
\end{figure}

\noindent\textbf{\textit{Author Keyword Co-occurrence}}

We analyzed the main keywords selected by the authors, which represent the central themes of the publications. In Figure~\ref{fig:network}, the co-occurrence of these keywords is visualized through a network graph, a widely used method in bibliometric analysis~\cite{lee2010investigating,kim2021machine}. Each node represents a keyword, and the edges between nodes indicate the co-occurrence of those keywords within individual publications. To improve clarity, edges representing fewer than three co-occurrences were removed after constructing the network graph.
Based on the most frequent occurrences of specific keywords, we categorized the main areas of research focus as follows: (i) `Mental Health,' (ii) `Chatbot,' (iii) `Artificial Intelligence,' and (iv) `Natural Language Processing (NLP).' 
Focusing on these keywords, we identified additional prominent terms.
In the context of `Mental health,' `sentiment analysis' was the most frequently cited keyword, followed by `anxiety,' `NLP,' and `dialogue system.' For `Chatbot'-related terms, `machine learning' was the most commonly used, followed by `depression,' `NLP,' `deep learning,' `anxiety,' and `dialogue system.'
Given that major trends can fluctuate annually, we also analyzed the author keywords on a year-by-year basis. Accordingly, we generated annual co-occurrence graphs for author keywords that appeared more than twice, as illustrated in Figure~\ref{fig:year_keyword}.
Firstly, in 2020, only 9 papers were published, leading to a relatively small number of overall keywords. The primary keywords were concentrated around `natural language processing,' `machine learning,', `chatbot' and `depression,' all of which were related to dialogue systems.
In 2021, prominent keywords related to mental health, such as `depression,' `anxiety,' and `COVID-19,' as well as chatbot technologies like `conversational agent,' `dialogue system,' and `conversational AI,' frequently emerged as central themes.
Next, in 2022, keywords centered around artificial intelligence began to appear, including `empathy,' `motivational interviewing,' and `virtual assistant.'
In 2023, which had the highest number of publications, a diverse range of keywords was identified. Empathy-related terms, such as `empathetic dialogue generation' and `empathetic response,' were particularly notable. Additionally, keywords like `emotional support conversation,' `motivational interviewing,' and `conversational robot' were also distinguished. This suggests that research on the mental health dialogue system became more specialized and diverse in 2023.
Interestingly, keywords like `large language model' and `GPT-3' appeared in 2024, reflecting a growing trend in the use of LLMs in mental health dialogue systems. There has also been an increase in keywords such as `emotional support' and `emotional support conversation,' as described in Figure~\ref{fig:year_keyword}. 

%% file: 4-experiment2-top10.tex
\subsection{Pre-LLM Approaches: Task-Specific Deep Learning Models}~\label{sec:pre_llm_qual}

\subsubsection{Overview of Highly Cited Top 10 Publications }~\label{sec:top10}

As presented in Table~\ref{tab:citation_table}, Scopus and WoS reported over 800 and 60 annual citations of published papers, respectively (as of May 7, 2024). Aligned with the increasing publication trends shown in Table~\ref{tab:year_table}, the annual citation count has also demonstrated a consistent upward trend. Note that ACM publications were excluded due to the inaccessibility of citation count data.
We then performed a qualitative analysis of the ten most frequently cited studies employing task-specific deep learning models.
In particular, we examined (i) the psychological perspective, focusing on aspects such as the ultimate goals for mental health improvement and the psychological approaches integrated into deep learning models. Additionally, we evaluated (ii) computational strategies, including the datasets utilized in the development of dialogue systems, the methodologies and theoretical frameworks applied in counseling systems, and the evaluation metrics used to assess their effectiveness. Table~\ref{tab:app_highlycited_table} provides a comprehensive summary of the publications.

\begin{table}[h!]
\centering
\resizebox{0.22\textwidth}{!}{%
\begin{tabular}{c|c|c}
\hline
\textbf{Year} &  \textbf{SCOPUS} &  \textbf{WOS} \\
\hline
2020 &  30 & 22 \\
2021 &  177 & 63 \\
2022 &  378 & 88 \\
2023 &  722 & 90 \\
2024 &  806 & 68 \\ \hline
Total&  2,113 & 331 \\ \hline
\end{tabular}
}
\caption{Number of Publication Citations per Year in Scopus and WoS}
\label{tab:citation_table}
\end{table}

\input{tab_highly10}


\noindent\textbf{\textit{Psychological Background}}

In this section, we reviewed psychological approaches to developing mental health dialogue systems.
\citet{kim2021perspective}, for example, aimed to improve empathetic engagement in dialogue systems by incorporating \textit{perspective-taking}, a psychological process that facilitates understanding situations from others' viewpoints. By employing the Rational Speech Acts (RSA) framework~\cite{frank2012predicting}, a probabilistic model that views communication as a recursive reasoning process, they utilized Bayesian network inferences with target words tied to emotional causes. This iterative approach allowed for inferring intentions and beliefs, enhancing the system's capacity to produce empathetic responses.
\citet{majumder-etal-2020-mime} also proposed a model that generates empathetic responses based on whether the emotions in prior responses are positive or negative, building on the idea that empathy involves mirroring another person's emotions~\cite{carr2003neural}.
Highlighting the importance of accurately understanding another person's circumstances and emotional states for effective empathy, \citet{sabour2022cem} employed Commonsense Transformers (COMET)~\cite{bosselut2019comet}, which leverage a commonsense knowledge graph to produce contextually relevant inferences about events, actions, and emotional states. This approach facilitates a more precise and nuanced comprehension of human behavior and interactions, enabling dialogue systems to respond with greater empathy and contextual awareness.
Moreover, \citet{sharma2021towards} introduced a reinforcement learning model that rewards performance based on fluency and consistency, which are critical factors in enhancing empathetic communication.
Both \citet{li-etal-2020-empdg} and \citet{shin2020generating} developed models that integrate feedback or sentiment from user responses to improve empathetic interactions, highlighting the importance of tracking client responses in fostering successful empathy.
In addition, drawing on empirical evidence that social media platform users can be categorized as information-seeking or emotion-seeking types, \citet{wang2021cass} first classified user types using a CNN module and provided empathetic responses only to emotion-seeking users.
\citet{liu2021towards} emphasized that counselors deliver emotional support by using psychological counseling strategies based on the context and information shared by clients. Consequently, the researchers designed the decoder to produce a specialized strategy token aligned with the prior conversational context, followed by generating a response conditioned on this token. Human evaluation outcomes indicated that this method significantly improved the model's emotional support quality, particularly in the areas of Fluency, Identification, Suggestion, and Comforting.

\noindent\textbf{\textit{Computational Approach}}

\noindent\textbf{Training Datasets:}
We observed that the main objective of dialogue systems for mental health improvement is influenced by the choice of dataset.
To be specific, our analysis revealed that numerous studies employed the publicly available EmpatheticDialogue dataset~\cite{rashkin2019towards}, comprising 25,000 conversations designed to address the emotional cues of dialogue partners.
On the other hand, some studies constructed their own datasets for training deep learning models. For instance, \citet{wang2021cass} collected post-response pairs from a pregnancy healthcare community developed to provide informational and emotional support to pregnant women. Likewise, \citet{sharma2021towards} collected posts and responses from TalkLife, an online peer-to-peer support platform, which were then refined by human experts to improve empathetic quality, converting interactions with lower empathy into those with higher empathy.
Although only one of the ten studies leveraged the ESConv dataset~\cite{liu2021towards}, we will delve into the ESConv dataset and related studies separately in a subsequent section. This dataset has recently garnered attention for its applicability, as its dialogue sets are annotated according to psychological counseling processes~\cite{hill2020helping}.

\noindent\textbf{Deep Learning Techniques:}
We noticed that most studies aimed to develop models for generating empathetic responses and built their models on the decoder architecture of transformers~\cite{vaswani2017attention}. 
Furthermore, two studies employed reinforcement learning models that utilized sentiment intensity~\cite{shin2020generating} and fluency and coherence~\cite{sharma2021towards} as reward signals to promote the generation of empathetic responses.
Remarkably, external knowledge sources were utilized to fill gaps in domain knowledge. For example, \citet{li2022knowledge} incorporated NRC-VAD~\cite{bosselut2019comet} to support the understanding of emotional tone; this resource provides human ratings for over 20,000 English words in terms of valence, arousal, and dominance. Similarly, \citet{sabour2022cem} used COMET~\cite{bosselut2019comet}, a commonsense knowledge graph designed to generate inferences aligned with the context of events, actions, and emotional states, to enhance the interpretation of emotions and contexts.
\citet{kim2021perspective} adopted Bayesian Inference to adjust prior beliefs based on observed data, such as emotion-inducing words, enabling the model to dynamically enhance its understanding of emotional triggers. This approach improved the contextual relevance and specificity of the empathetic responses generated.
Owing to the recent advancements in techniques, publications utilizing LLMs in dialogue systems were not included among the top 10 papers. However, given their strong performance in language generation, we will discuss LLMs and related studies individually.

\input{tab_evaluation}

\noindent\textbf{Evaluation Metrics:}
We found that there is a wide range of metrics used to evaluate the performance of empathetic response generation, with no standardized approach. The description of each metric is provided in Table~\ref{tab:app_combined_metrics}. 
Specifically, many studies utilized statistical automatic metrics such as ACC~\cite{li-etal-2020-empdg,sabour2022cem,li2022knowledge}, DIST~\cite{li-etal-2020-empdg,sharma2021towards,sabour2022cem,li2022knowledge,shin2020generating}, BLEU~\cite{majumder-etal-2020-mime,liu2021towards,wang2021cass,shin2020generating,lin2020caire}, and PPL~\cite{li-etal-2020-empdg,sharma2021towards,liu2021towards,sabour2022cem,li2022knowledge,lin2020caire}. Moreover, by utilizing pretrained language models, researchers have been able to evaluate a broader range of factors. For instance, \citet{shin2020generating} employed Bag of Word Embedding Similarity~\cite{liu2016not} to assess inter-sentence similarity, while \citet{kim2021perspective} used a trained RoBERTa model~\cite{liu2019roberta} to measure counseling strategies, such as exploration and interpretation, to capture empathetic attributes.
Nevertheless, automatic metrics like BLEU have been found to correlate weakly with human assessments of response quality~\cite{liu2016not}. Thus, human evaluation methods were adopted to measure complex empathetic responses more accurately~\cite{majumder-etal-2020-mime}. Annotators evaluated general responses across dimensions such as `Empathy,' `Relevance,' and `Fluency'~\cite{kim2021perspective,li-etal-2020-empdg,majumder-etal-2020-mime,li2022knowledge}. In a similar effort, \citet{wang2021cass} assessed human and AI responses based on human judgments concerning `Grammar Correctness,' `Relevance,' `Willingness to Reply,' and `Emotional Support.' A/B testing was also frequently employed to determine which model appeared more empathetic and human-like~\cite{kim2021perspective,li-etal-2020-empdg,majumder-etal-2020-mime,liu2021towards}.

%% file: tab_highly10.tex
\begin{table}[!ht]
\centering
\resizebox{\columnwidth}{!}{%
\begin{tabular}{@{}c|c|c|c|c|c|c|c|c@{}}
\toprule
\textbf{Reference} & \textbf{Year} & \textbf{Task} & \textbf{\begin{tabular}[c]{@{}c@{}}Existing\\ Gaps\end{tabular}} & \textbf{Strategy} & \textbf{\begin{tabular}[c]{@{}c@{}}Psychological\\ Background\end{tabular}} & \textbf{\begin{tabular}[c]{@{}c@{}}Computational\\  Approach\end{tabular}} & \textbf{\begin{tabular}[c]{@{}c@{}}Training\\ Dataset\end{tabular}} & \textbf{\begin{tabular}[c]{@{}c@{}}Evaluation\\ Metrics\end{tabular}} \\ \midrule
\begin{tabular}[c]{@{}c@{}}Majumder\\  et al. ~\cite{majumder-etal-2020-mime}\end{tabular} & 2020 & ERG & \begin{tabular}[c]{@{}c@{}}Ambivalent response \\ to a positive utterance\end{tabular} & \begin{tabular}[c]{@{}c@{}}Mimic the emotion \\ of the speaker\end{tabular} & \begin{tabular}[c]{@{}c@{}}Empathy involves\\ reflecting others’ \\ emotions ~\cite{carr2003neural}\end{tabular} & \begin{tabular}[c]{@{}c@{}}Emotion stochastic \\ sampling, \\ Emotion mimicry\end{tabular} & \begin{tabular}[c]{@{}c@{}}Empathetic \\  Dialogue~\cite{rashkin2019towards}\end{tabular} & \begin{tabular}[c]{@{}c@{}}BLEU, Empathy, \\ Relevance, Fluency\end{tabular} \\ \hline
\begin{tabular}[c]{@{}c@{}}Li et al.\\  ~\cite{li-etal-2020-empdg}\end{tabular} & 2020 & ERG & \begin{tabular}[c]{@{}c@{}}Lack of potential \\ user feedback use\end{tabular} & \begin{tabular}[c]{@{}c@{}}Feedback-based \\ response generation\end{tabular} & \begin{tabular}[c]{@{}c@{}}Response conveys \\ feedback on prior \\ utterances ~\cite{zhang2018exploring}\end{tabular} & \begin{tabular}[c]{@{}c@{}}Adversarial Learning \\ (1) Generator\\  (2) Emotional/Semantic \\ Discriminator\end{tabular} & \begin{tabular}[c]{@{}c@{}}Empathetic \\  Dialogue~\cite{rashkin2019towards}\end{tabular} & \begin{tabular}[c]{@{}c@{}}ACC, PPL,\\  DIST-1, DIST-2, \\ Empathy, Relevance, \\ Fluency\end{tabular} \\ \hline
\begin{tabular}[c]{@{}c@{}}Lin et al.\\  ~\cite{lin2020caire}\end{tabular} & 2020 & ERG & \begin{tabular}[c]{@{}c@{}}Independent, complex \\ system design\end{tabular} & End-to-end Model & - & \begin{tabular}[c]{@{}c@{}}NLU-NLG \\ integration\end{tabular} & \begin{tabular}[c]{@{}c@{}}Empathetic \\  Dialogue~\cite{rashkin2019towards},\\  PersonaChat ~\cite{zhang-etal-2018-personalizing}\end{tabular} & \begin{tabular}[c]{@{}c@{}}PPL, AVG-BLEU,\\  EMO-ACC\end{tabular} \\ \hline
\begin{tabular}[c]{@{}c@{}}Sharma et\\  al. ~\cite{sharma2021towards}\end{tabular} & 2021 & \begin{tabular}[c]{@{}c@{}}Empathetic \\ Rewriting\end{tabular} & \begin{tabular}[c]{@{}c@{}}Lack of scalability \\ for online platforms\end{tabular} & \begin{tabular}[c]{@{}c@{}}Empathetic response \\ suggestion \\ for supporters\end{tabular} & \begin{tabular}[c]{@{}c@{}}Empathetic interactions \\ contribute to improving \\ mental health ~\cite{elliott2018therapist}.\end{tabular} & \begin{tabular}[c]{@{}c@{}}RL w/ empathy\\ improvement rewards\end{tabular} & \begin{tabular}[c]{@{}c@{}}TALKTOME\\  community\end{tabular} & \begin{tabular}[c]{@{}c@{}}Change in empathy, \\ PPL, Coherence, \\ Extrema, DIST-1, \\ DIST-2, Edit rate\end{tabular} \\ \hline
\begin{tabular}[c]{@{}c@{}}Liu et al.\\  ~\cite{liu2021towards}\end{tabular} & 2021 & ESC & \begin{tabular}[c]{@{}c@{}}Lack of understanding\\ help-seeker context\end{tabular} & \begin{tabular}[c]{@{}c@{}}Emotional support \\ through strategy \\ selection ~\cite{hill2020helping}\end{tabular} & Helping Skills Theory ~\cite{hill2020helping} & Strategy token prepending & ESConv ~\cite{liu2021towards} & \begin{tabular}[c]{@{}c@{}}BLEU-2, \\ ROUGE-L, Extrema\end{tabular} \\ \hline
\begin{tabular}[c]{@{}c@{}}Sabour et\\  al. ~\cite{sabour2022cem}\end{tabular} & 2022 & ERG & \begin{tabular}[c]{@{}c@{}}Insufficient \\ cognitive empathy\end{tabular} & \begin{tabular}[c]{@{}c@{}}Capture \\ situational context\end{tabular} & \begin{tabular}[c]{@{}c@{}}Empathy includes \\ affective and cognitive \\ components ~\cite{davis1980multidimensional}\end{tabular} & \begin{tabular}[c]{@{}c@{}}Cognitive context \\ via External knowledge \\ (COMET)\end{tabular} & \begin{tabular}[c]{@{}c@{}}Empathetic \\ Dialogue~\cite{rashkin2019towards}\end{tabular} & \begin{tabular}[c]{@{}c@{}}ACC, PPL,\\  DIST-1, DIST-2\end{tabular} \\ \hline
\begin{tabular}[c]{@{}c@{}}Li et al.\\  ~\cite{li2022knowledge}\end{tabular} & 2022 & ERG & \begin{tabular}[c]{@{}c@{}}Lack of understanding\\ implicit emotion\end{tabular} & \begin{tabular}[c]{@{}c@{}}Detect implicit \\ emotion cues\end{tabular} & - & \begin{tabular}[c]{@{}c@{}}External knowledge\\  (NRC-VAD, ConceptNet),\\  GNN with attention\end{tabular} & \begin{tabular}[c]{@{}c@{}}Empathetic \\ Dialogue~\cite{rashkin2019towards}\end{tabular} & \begin{tabular}[c]{@{}c@{}}ACC, PPL,\\ DIST-1, DIST-2, \\ Empathy, Relevance, \\ Fluency\end{tabular} \\ \hline
\begin{tabular}[c]{@{}c@{}}Wang et\\  al. ~\cite{wang2021cass}\end{tabular} & 2021 & ESC & \begin{tabular}[c]{@{}c@{}}Limited scalability \\ of rule-based systems\end{tabular} & \begin{tabular}[c]{@{}c@{}}Online \\ Community-wide\\ empathetic support\end{tabular} & - & \begin{tabular}[c]{@{}c@{}}CNN,\\  seq2seq with attention\end{tabular} & \begin{tabular}[c]{@{}c@{}}YouBao health \\ community posts\end{tabular} & \begin{tabular}[c]{@{}c@{}}ACC, BLEU,\\  Grammar, Relevance, \\ Correctness, \\ Willing-to-reply, \\ Emotional Support\end{tabular} \\ \hline
\begin{tabular}[c]{@{}c@{}}Kim et al.\\  ~\cite{kim2021perspective}\end{tabular} & 2021 & \begin{tabular}[c]{@{}c@{}}ERG, \\ Emotion \\ Recognition\end{tabular} & \begin{tabular}[c]{@{}c@{}}Generic  Expressions \\ show weak empathy\end{tabular} & \begin{tabular}[c]{@{}c@{}}Detect emotion word \\ and reflect on response\end{tabular} & \begin{tabular}[c]{@{}c@{}}Perspective-taking \\ is essential for \\ empathetic reasoning ~\cite{decety2005perspective}.\end{tabular} & Baysian Inference & \begin{tabular}[c]{@{}c@{}}Empathetic \\ Dialogue ~\cite{rashkin2019towards}\end{tabular} & \begin{tabular}[c]{@{}c@{}}Recall, EPITOME, \\ Empathy, Relevance, \\ Fluency\end{tabular} \\ \hline
\begin{tabular}[c]{@{}c@{}}Shin et al.\\ ~\cite{shin2020generating}\end{tabular} & 2020 & ERG & \begin{tabular}[c]{@{}c@{}}No consideration of \\ user’s future emotion reaction\end{tabular} & Predict future emotion & - & \begin{tabular}[c]{@{}c@{}}RL w/ future \\ sentiment improvement \\ reward\end{tabular} & \begin{tabular}[c]{@{}c@{}}Empathetic \\ Dialogue ~\cite{rashkin2019towards}\end{tabular} & \begin{tabular}[c]{@{}c@{}}DIST-1, DIST-2, \\ DIST-3, Extrema,\\ AVG-BLEU\end{tabular} \\ \bottomrule
\end{tabular}%
}
\caption{Overview of research methodologies utilized in highly cited publications. 
\\\textbf{ERG}: Empathetic Response Generation 
\\\textbf{ESC}: Emotional Support Conversation
\\\textbf{NLU}: Natural Language Understanding
\\\textbf{NLG}: Natural Language Generation
\\\textbf{RL}: Reinforcement Learning
}
\label{tab:app_highlycited_table}
\end{table}

%% file: tab_evaluation.tex
\begin{table}[tbp]
\centering
\resizebox{\columnwidth}{!}{%
\begin{tabular}{ccclc}
\hline
\textbf{Category} & \textbf{Metric} & \textbf{Type} & \textbf{Description} & \textbf{Reference} \\ \hline
\multirow{4}{*}{\begin{tabular}[c]{@{}c@{}}Performance\\ \& Accuracy\end{tabular}} & ACC & Statistical & Accuracy – percentage of correct predictions. & ~\cite{li-etal-2020-empdg, li2022knowledge, peng2023fado, sabour2022cem, tu2022misc, wang2021cass, zhao2023transesc} \\
 & INTENTACC & Model-based & \begin{tabular}[c]{@{}l@{}}Measures response intent accuracy via BERT fine-tuned \\ on EMPINTENT dataset ~\cite{welivita2020taxonomy}\end{tabular} & ~\cite{lee2022does} \\
 & EMOACC & Model-based & Measures emotion accuracy for empathetic response generation & ~\cite{lee2022does, lin2020caire} \\
 & PPL & Statistical & Perplexity – how well the model predicts a sequence; lower is better. & ~\cite{cheng2022improving, deng2023knowledge, lai2023supporting, lee2022does, li-etal-2020-empdg, li2022knowledge, lin2020caire, peng2022ijcai, peng2023fado, sabour2022cem, sharma2021towards, tu2022misc, zhao2023transesc, zhou2023facilitating} \\ \hline
\multirow{7}{*}{\begin{tabular}[c]{@{}c@{}}Text Quality\\ \& Diversity\end{tabular}} & BLEU-N & Statistical & Measures text similarity using n-gram overlap (N is the n-gram length). & ~\cite{chen2023emotional, chen2023soulchat, cheng2022improving, deng2023knowledge, firdaus2023multi, kaysar2023mental, lin2020caire, liu2021towards, majumder-etal-2020-mime, peng2022ijcai, peng2023fado, qian2023harnessing, shin2020generating, tu2022misc, wang2021cass, zhang-etal-2023-ask, zhao2023transesc, zhou2023facilitating} \\
 & ROUGE-L & Statistical & \begin{tabular}[c]{@{}l@{}}Evaluates longest common subsequence between texts \\ (L denotes the longest common subsequence)\end{tabular} & ~\cite{chen2023soulchat, cheng2022improving, deng2023knowledge, firdaus2023multi, kaysar2023mental, lai2023supporting, liu2021towards, peng2022ijcai, peng2023fado, tu2022misc, zhang-etal-2023-ask, zhao2023transesc} \\
 & METEOR & Statistical & Evaluate accuracy, fidelity, word order, and lexical diversity & ~\cite{cheng2022improving, peng2023fado, tu2022misc} \\
 & DIST-N & Statistical & Diversity – counts unique n-grams to assess variation in responses. & ~\cite{chen2023emotional, lai2023supporting, lee2022does, li-etal-2020-empdg, li2022knowledge, peng2022ijcai, peng2023fado, qian2023harnessing, sabour2022cem, sharma2021towards, shin2020generating, tu2022misc, zhao2023transesc, zhou2023facilitating} \\
 & Extrema & Model-based & \begin{tabular}[c]{@{}l@{}}Measrues semantic similarity by comparing the highest word embedding \\ values between generated and reference\end{tabular} & ~\cite{liu2021towards, sharma2021towards, shin2020generating} \\
 & BERTScore & Model-based & \begin{tabular}[c]{@{}l@{}}Evaluates semantic similarity between predicted and reference sentences \\ using token embeddings. P,R,F means Precision, Recall , F1 respectively.\end{tabular} & ~\cite{qian2023harnessing, zhang-etal-2023-ask} \\
 & BARTScore & Model-based & \begin{tabular}[c]{@{}l@{}}Evaluates generation quality using a pretrained BART model to score \\ the likelihood of generated text given the reference.\end{tabular} & ~\cite{zhang-etal-2023-ask} \\ \hline
\multirow{5}{*}{\begin{tabular}[c]{@{}c@{}}Coherence\\ \& Relevance\end{tabular}} & cDC & Model-based & Context Dialogue Coherence – fit within current conversation context & ~\cite{zhou2023facilitating} \\
 & fDC & Model-based & Future Dialogue Coherence – coherence with future user utterance & ~\cite{zhou2023facilitating} \\
 & Coherence & Statistical & Measures logical flow and structure of rewritten responses & ~\cite{sharma2021towards} \\
 & Relevance & Human & \begin{tabular}[c]{@{}l@{}}Measures the degree to which a response is contextually aligned \\ with the thematic content of the preceding dialogue\end{tabular} & ~\cite{kim2021perspective, li-etal-2020-empdg, li2022knowledge, majumder-etal-2020-mime, wang2021cass} \\
 & Edit Rate & Statistical & Measures how much a response was rewritten (precision/conciseness) & ~\cite{sharma2021towards} \\ \hline
\multirow{9}{*}{\begin{tabular}[c]{@{}c@{}}Emotional,\\ Empathetic Response\end{tabular}} & \begin{tabular}[c]{@{}c@{}}Change \\ in Empathy\end{tabular} & Statistical & \begin{tabular}[c]{@{}l@{}}Measures the increase or decrease in empathy levels in rewritten \\ responses compared to the original\end{tabular} & ~\cite{sharma2021towards} \\
 & \begin{tabular}[c]{@{}c@{}}EPITOME \\ (Emotion)\end{tabular} & Model-based & Captures warmth, compassion, and concern in response & ~\cite{kim2021perspective} \\
 & \begin{tabular}[c]{@{}c@{}}EPITOME \\ (Exploration)\end{tabular} & Model-based & Measures ability to explore and understand user emotions & ~\cite{kim2021perspective} \\
 & \begin{tabular}[c]{@{}c@{}}EPITOME \\ (Interpretation)\end{tabular} & Model-based & Evaluates if the model understands inferred user experiences & ~\cite{kim2021perspective} \\
 & cES & Model-based & \begin{tabular}[c]{@{}l@{}}Conversation-level Emotional Support \\ – Measures the percentage of correct predictions\end{tabular} & ~\cite{zhou2023facilitating} \\
 & tES & Model-based & Turn-level Emotional Support – Prediction accuracy for each turn & ~\cite{zhou2023facilitating} \\
 & Valence & Human & \begin{tabular}[c]{@{}l@{}}Evaluates the positivity or negativity of the emotional response elicited \\ by the agent, reflecting the overall emotional tone\end{tabular} & ~\cite{llanes2024developing} \\
 & Arousal & Human & \begin{tabular}[c]{@{}l@{}}Measures the level of physiological and emotional activation elicited \\ by the agent, ranging from calm to excited states\end{tabular} & ~\cite{llanes2024developing} \\
 & Empathy & Human & \begin{tabular}[c]{@{}l@{}}Which model showed more suitable emotional responses, \\ like warmth and concern?\end{tabular} & ~\cite{chen2023emotional, chen2023soulchat, li-etal-2020-empdg, li2022knowledge, majumder-etal-2020-mime, qian2023harnessing, tu2022misc, zhao2023transesc} \\ \hline
\multirow{5}{*}{\begin{tabular}[c]{@{}c@{}}Human-likeness\\ \& Naturalness\end{tabular}} & Fluency & Human & Which bot’s responses were clearer and more natural? & ~\cite{cheng2022improving, deng2023knowledge, kim2021perspective, li-etal-2020-empdg, li2022knowledge, majumder-etal-2020-mime, peng2022ijcai, peng2023fado, qian2023harnessing, tu2022misc, zhao2023transesc, zhou2023facilitating} \\
 & Naturalness & Human & Measures how smooth and natural the model's response feels & ~\cite{chen2023soulchat, llanes2024developing} \\
 & Realism & Human & \begin{tabular}[c]{@{}l@{}}Measures how closely the agent interaction aligns with authentic \\ human conversation characteristics\end{tabular} & ~\cite{llanes2024developing} \\
 & Rapport & Human & \begin{tabular}[c]{@{}l@{}}Measures the connection established between clients and \\ counselors in empathetic dialogs\end{tabular} & ~\cite{chen2023emotional} \\
 & Coherence & Human & Which bot’s response better fit the context across turns? & ~\cite{chen2023emotional, kharitonova2024incorporating, qian2023harnessing, zhou2023facilitating} \\ \hline
\multirow{5}{*}{\begin{tabular}[c]{@{}c@{}}Informativeness\\ \& Safety\end{tabular}} & Informativeness & Human & Which bot’s response was more varied, detailed, and informative? & ~\cite{zhou2023facilitating} \\
 & Helpfulness & Human & Assesses the usefulness of the model's response to the user's needs & ~\cite{chen2023emotional, chen2023soulchat} \\
 & Varacity & Human & \begin{tabular}[c]{@{}l@{}}Measures whether the response generated by the system is correct \\ and relevant to the question\end{tabular} & ~\cite{kharitonova2024incorporating} \\
 & Evidence & Human & \begin{tabular}[c]{@{}l@{}}Measures whether response includes references to relevant \\ studies, clinical trials, or guidelines that can validate it\end{tabular} & ~\cite{kharitonova2024incorporating} \\
 & Safety & Human & \begin{tabular}[c]{@{}l@{}}Measures whether the model's response avoids harmful, \\ offensive, or legally sensitive content\end{tabular} & ~\cite{chen2023soulchat} \\ \hline
\multirow{5}{*}{\begin{tabular}[c]{@{}c@{}}User-Centered \\ Evaluation\end{tabular}} & Identification & Human & Which bot better analyzed your situation and identified your problems? & ~\cite{cheng2022improving, deng2023knowledge, peng2022ijcai, peng2023fado, qian2023harnessing, zhao2023transesc} \\
 & Comforting & Human & Which bot was more effective in providing comfort? & ~\cite{cheng2022improving, deng2023knowledge, peng2022ijcai, peng2023fado} \\
 & Suggestion & Human & Which model offered more useful advice? & ~\cite{cheng2022improving, deng2023knowledge, peng2022ijcai, peng2023fado, zhao2023transesc} \\
 & Supportiveness & Human & Which bot was more effective at shifting the user’s emotions positively? & ~\cite{zhou2023facilitating} \\
 & Knowledge & Human & Extent to which useful knowledge is provided & ~\cite{tu2022misc} \\ \hline
\end{tabular}%
}
\caption{Explanation of Evaluation Metrics for Automatic and Human Assessment}
\label{tab:app_combined_metrics}
\end{table}

%% file: 4-experiment3-ESconv.tex
\subsubsection{Overview of Publications using Emotional Support Conversation (ESConv) Dataset}~\label{sec:esconv}

\begin{figure}[!ht]
    \centering
    \includegraphics[width=0.9\textwidth]{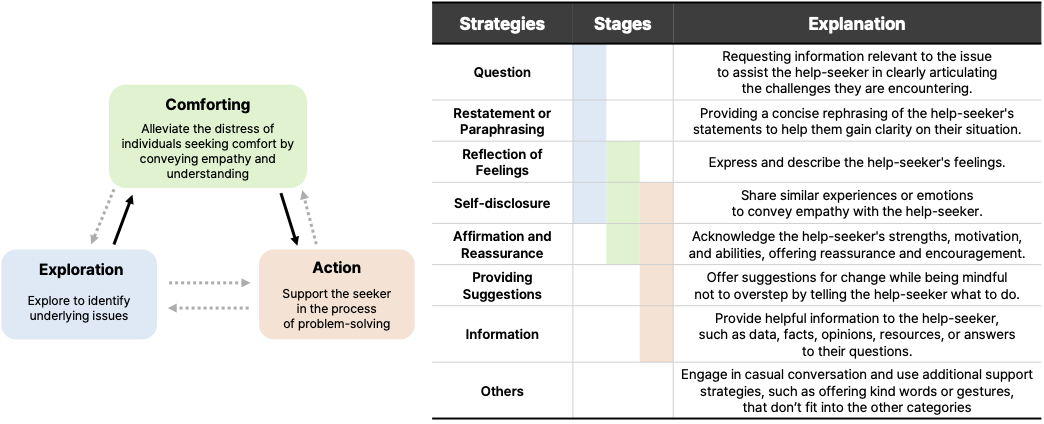}
    \caption{We present figures from the original paper \cite{liu2021towards} to illustrate three stages of the ESConv framework and the corresponding eight counseling strategies. According to \citet{liu2021towards}, this framework comprises three stages, each with specific support strategies. The \textit{exploration} stage aims to help individuals identify underlying issues; the \textit{comforting} stage focuses on providing empathy and understanding; and the \textit{action} stage involves offering practical information or suggestions. Typically, the emotional support process follows a sequential order from 1. Exploration → 2. Comforting → 3. Action, as indicated by black arrows, though it can also be adjusted to suit the conversation's needs, as represented by dashed gray arrows.}
    \label{fig:esconv_stage}
\end{figure}

Providing emotional support is an essential skill in mental health interventions, aiming to alleviate emotional distress and assist individuals in navigating the challenges they encounter~\cite{langford1997social}.
In line with this, \citet{liu2021towards} developed the Emotional Support Conversation (ESConv) dataset, which consists of dialogues between trained crowd workers acting as supporters and help-seekers, who were required to complete a pre-chat survey about their problems and emotions, as well as provide feedback during and after the conversations. As described in Figure~\ref{fig:esconv_stage}, each conversational turn in ESConv is annotated by selecting the appropriate strategy and corresponding stage in the counseling process based on principles from Helping Skills Theory~\cite{hill2020helping}. This approach mirrors that of trained psychological counselors, who select strategies and stages based on the context of the conversation~\cite{hill2020helping}.
Moreover, the dataset includes the help-seeker's final emotional intensity, as well as assessments of the supporters' empathy and the relevance of their responses following each conversation.

The ESConv dataset is highly aligned with psychological counseling processes and has been widely cited, with 330 citations on Google Scholar (Accessed on June 27, 2025). Recognizing its importance, we conducted a comprehensive review of eight studies that leveraged this dataset, highlighting at least one citation from the selected studies. The summary of selected papers is described in Table~\ref{tab:app_esconv}.

\input{tab_esconv}
 
\noindent\textbf{\textit{Strategies to Address Existing Research Gaps}}

We discovered that, despite using the same dataset, selected studies adopt differing psychological perspectives on improving emotional support, which has resulted in the development of diverse approaches.
Numerous studies highlighted the importance of understanding an individual's psychological state for delivering effective emotional support.
For instance, \citet{tu2022misc} enhanced comprehension of help-seeker's emotions by utilizing the commonsense knowledge graph dataset, COMET~\cite{bosselut2019comet}.
\citet{peng2022ijcai} developed a hierarchical graph network to capture both the seeker's overarching concerns throughout the conversation and the peripheral intentions within each utterance, suggesting that emotional support can be enhanced by understanding the root of the problem.
Similarly, \citet{zhou2023facilitating} highlighted that the importance of emotion elicitation for better emotiaon support, leading them to use the emotion intensity factor in ESConv as the reward function within the Reinforcement Learning framework.
Conversely, a few studies emphasized the significance of selecting appropriate strategies to improve the emotional satisfaction of help-seekers. 
\citet{liu2021towards} proposed that strategic decision-making plays a crucial role in providing emotional support, demonstrating the effectiveness of identifying anticipated strategies in generating empathetic responses. \citet{peng2023fado} incorporated user feedback into strategy selection by reinforcing or discouraging specific counseling strategies based on the help-seeker's responses throughout the conversation, resulting in more accurate and user-aligned strategy predictions. 
Furthermore, \citet{cheng2022improving} underscored the importance of long-term strategy planning, prompting them to forecast future user feedback and optimize emotional support strategies across multiple conversational turns.
Although the ESConv dataset has shown promise in supporting the development of counseling dialogue systems for emotional support, it remains insufficient for capturing real-world scenarios and the full range of counseling strategies necessary for developing comprehensive psychological counseling systems. We will address the limitations of datasets for counseling system development in the discussion section.

\noindent\textbf{\textit{Computational Approach}}

\noindent\textbf{Deep Learning Techniques:}
We identified two main approaches commonly utilized in the selected papers: (i) multi-task learning and (ii) the integration of external knowledge resources.
First, several studies adopted multi-task learning, wherein a single model is trained on multiple tasks concurrently to enhance performance through shared representations. For example, \citet{peng2022ijcai} enabled the model to simultaneously generate responses and predict the type of issue presented by the help-seeker, while other studies allowed the model to generate responses and identify the counseling strategies applied by the supporter~\cite{liu2021towards,tu2022misc,peng2023fado}. Additionally, the joint prediction of keywords and emotions was also investigated~\cite{zhou2023facilitating,zhao2023transesc}.
As related tasks can significantly improve model performance by facilitating knowledge transfer between tasks~\cite{fifty2021efficiently}, we noticed that many studies leveraged the training of similar tasks together to maximize these benefits.
Second, external knowledge sources have also been incorporated to address the limitations of domain knowledge in training data and models. For instance, \citet{deng2023knowledge} employed HEAL~\cite{welivita2022heal}, a knowledge graph specifically designed for mental health conversations. COMET~\cite{bosselut2019comet} also has been utilized to infer users' emotions~\cite{tu2022misc,zhou2023facilitating,zhao2023transesc}, intentions and underlying psychological causes~\cite{peng2022ijcai}. In addition, NRC-VAD~\cite{mohammad-2018-obtaining} has been used for emotion detection~\cite{cheng2022improving,zhou2023facilitating}. 
Furthermore, we examined the underlying backbone models adapted to implement those approaches. \citet{zhao2023transesc} introduced a State Transition Graph (STG)~\cite{naldi2011dynamically} to represent the dynamic behavior of directed graphs. Using this approach, they tracked semantic, emotional, and strategic transitions throughout conversations by constructing a separate graph for each stage of counseling. Likewise, \citet{peng2022ijcai} built a hierarchical graph attention network to model the relationships among the global cause, local intention, and dialogue history.
\citet{tu2022misc} suggested a strategy probability distribution method to select subsequent counseling strategies, mapping the probability of each strategy's selection to a discrete latent space. This approach allows the model to consider multiple strategies in a dynamic rather than fixed manner, facilitating the generation of responses that are both supportive and contextually relevant.
Furthermore, \citet{cheng2022improving} applied the A* algorithm for strategy planning, an optimal path-finding algorithm that calculates the shortest path using both cost and heuristic functions~\cite{hart1968formal}.
\citet{zhou2023facilitating}, in addition, implemented a Mixture of Experts (MoE) architecture for emotion and keyword predictions, combined with reinforcement learning that uses emotion intensity and coherence as reward signals. MoE is a neural network architecture designed to improve efficiency and performance by routing inputs to specialized sub-models~\cite{jordan1994hierarchical}. 
Finally, \citet{peng2023fado} managed the selection of subsequent strategies and context representation through a gating mechanism, a trainable component that regulates information flow by learning to selectively allow or block signals through multiplicative operations~\cite{gu2020improving}.

\noindent\textbf{Automatic Evaluation Metrics:}
A variety of statistical evaluation metrics have been applied to assess the performance of emotional support systems, including PPL~\cite{liu2021towards,peng2022ijcai,tu2022misc,peng2023fado,cheng2022improving,deng2023knowledge,zhou2023facilitating,zhao2023transesc}, BLEU~\cite{liu2021towards,peng2022ijcai,tu2022misc,peng2023fado,cheng2022improving,deng2023knowledge,zhou2023facilitating,zhao2023transesc}, ROUGE-L~\cite{liu2021towards,peng2022ijcai,tu2022misc,peng2023fado,cheng2022improving,deng2023knowledge,zhao2023transesc}, DIST~\cite{peng2022ijcai,tu2022misc,peng2023fado,zhou2023facilitating,zhao2023transesc}, METEOR~\cite{cheng2022improving}, CIDEr~\cite{cheng2022improving}, Extrema~\cite{liu2021towards}, and ACC~\cite{tu2022misc,peng2023fado,zhao2023transesc}. 
\citet{zhou2023facilitating} introduced evaluation metrics---cES and tES for assessing emotional elicitation intensity, and cDC and fDC for measuring response coherence. These metrics were evaluated using the pre-trained emotion classification model DistilRoBERTa~\cite{hartmann2022emotionenglish} along with BERT~\cite{devlin2019bert}.
Detailed explanations of these metrics can be found in Table~\ref{tab:app_combined_metrics}.
However, our findings indicate that the evaluation metrics predominantly assess fluency and accuracy, often overlooking the semantic richness of responses. These metrics are notably limited in their ability to capture semantic equivalence between sentences that differ in wording yet express similar meanings~\cite{lowe2017towards}.
To more accurately evaluate the semantic content of responses, it is essential to incorporate diverse and semantically-aware evaluation methods, such as BERTScore~\cite{bert-score}, which leverages cosine similarity to align sentence embeddings.

\noindent\textbf{Human Evaluation Metrics:}
To effectively assess emotional support conversations, it is essential to incorporate human evaluation~\cite{liu2021towards}, which may provide valuable insights into subtle elements such as empathy, rapport, and perceived helpfulness that automated systems may not fully capture.
Our analysis found that most publications considered human A/B evaluations, which compare responses from the target model against a baseline to determine which is superior. The factors evaluated to determine the quality of emotional support included  Suggestion~\cite{liu2021towards,peng2023fado,cheng2022improving,zhou2023facilitating,zhao2023transesc,deng2023knowledge,peng2022ijcai}, Identification~\cite{liu2021towards,peng2022ijcai,cheng2022improving,peng2023fado,zhao2023transesc,deng2023knowledge}, Empathy~\cite{cheng2022improving,zhao2023transesc}, Informativeness~\cite{zhou2023facilitating}, Coherence~\cite{zhou2023facilitating}, and Supportiveness~\cite{zhou2023facilitating}.
In contrast, \citet{tu2022misc} assembled three experts with backgrounds in linguistics or psychology to independently assess Fluency, Knowledge, and Empathy, rating each on a scale from 0 to 2.
Detailed descriptions of each evaluation aspect can be found in Table~\ref{tab:app_combined_metrics}. 

%% file: tab_esconv.tex
\begin{table}[!ht]
\centering
\resizebox{\columnwidth}{!}{%
\begin{tabular}{@{}c|c|c|c|c|c|c@{}}
\toprule
\textbf{Reference} & \textbf{Year} & \textbf{\begin{tabular}[c]{@{}c@{}}Existing\\ Gaps\end{tabular}} & \textbf{Strategy} & \textbf{\begin{tabular}[c]{@{}c@{}}Computational\\ Approach\end{tabular}} & \textbf{\begin{tabular}[c]{@{}c@{}}Automatic\\ Evaluation\end{tabular}} & \textbf{\begin{tabular}[c]{@{}c@{}}Human\\ Evaluation\end{tabular}} \\ \midrule
Liu et al.~\cite{liu2021towards} & 2021 & \begin{tabular}[c]{@{}c@{}}Lack of understanding\\ help-seeker context\end{tabular} & \begin{tabular}[c]{@{}c@{}}Emotional support\\ through \\ strategy selection\end{tabular} & \begin{tabular}[c]{@{}c@{}}Strategy token \\ prepending\end{tabular} & \begin{tabular}[c]{@{}c@{}}BLEU-2, \\ ROUGE-L, \\ Extrema\end{tabular} & \begin{tabular}[c]{@{}c@{}}Fluency, Identification, \\ Comforting, Suggestion\end{tabular} \\ \hline
Peng et al.~\cite{peng2022ijcai} & 2022 & \begin{tabular}[c]{@{}c@{}}Overlooks hierarchical \\ structure of causes \\ and psychological intent\end{tabular} & \begin{tabular}[c]{@{}c@{}}Detect implicit \\ emotional problem \\ and current speaker’s state\end{tabular} & \begin{tabular}[c]{@{}c@{}}External knowledge \\ (COMET),\\  Hierarchical Graph, \\ Attention Network\end{tabular} & \begin{tabular}[c]{@{}c@{}}PPL, BLEU-4, \\ DIST-2,\\  ROUGE-L\end{tabular} & \begin{tabular}[c]{@{}c@{}}Fluency, Identification, \\ Comforting, Suggestion\end{tabular} \\ \hline
Tu et al.~\cite{tu2022misc}& 2022 & \begin{tabular}[c]{@{}c@{}}Relies on coarse \\ emotion labels, \\ missing fine-grained \\ user state\end{tabular} & \begin{tabular}[c]{@{}c@{}}Enhance the comprehension \\ of the seeker’s emotion\end{tabular} & \begin{tabular}[c]{@{}c@{}}External knowledge\\  (COMET),\\  Cross-attention\end{tabular} & \begin{tabular}[c]{@{}c@{}}ACC, PPL,\\  DIST-2, BLEU-4, \\ ROUGE-L, METEOR\end{tabular} & \begin{tabular}[c]{@{}c@{}}Fluency, \\ Knowledge, \\ Empathy\end{tabular} \\ \hline
Peng et al. ~\cite{peng2023fado} & 2023 & \begin{tabular}[c]{@{}c@{}}Insufficient use \\ of user feedback \\ and strategy-to-context \\ flow\end{tabular} & \begin{tabular}[c]{@{}c@{}}Considering \\ seeker’s feedback \\ when selecting strategy\end{tabular} & \begin{tabular}[c]{@{}c@{}}Gate mechanism,\\  Strategy dictionary\end{tabular} & \begin{tabular}[c]{@{}c@{}}ACC, PPL,\\  DIST-2, BLEU-4, \\ ROUGE-L, METEOR\end{tabular} & \begin{tabular}[c]{@{}c@{}}Fluency, Identification, \\ Comforting, Suggestion\end{tabular} \\ \hline
Cheng et al. ~\cite{cheng2022improving} & 2022 & \begin{tabular}[c]{@{}c@{}}Over-simplified \\ single-turn \\ interaction assumption\end{tabular} & \begin{tabular}[c]{@{}c@{}}Long-term strategy\\ planning through \\ forecasting the future’s \\ seeker’s feedback\end{tabular} & \begin{tabular}[c]{@{}c@{}}Emotion cause detection, \\ External knowledge \\ (ERC-VAD), \\ A* algorithm\end{tabular} & \begin{tabular}[c]{@{}c@{}}PPL, BLEU-4, \\ ROUGE-L, METEOR, \\ CIDEr\end{tabular} & \begin{tabular}[c]{@{}c@{}}Fluency, Identification, \\ Comforting, Suggestion\end{tabular} \\ \hline
Deng et al.~\cite{deng2023knowledge}& 2023 & \begin{tabular}[c]{@{}c@{}}Lack of proactive \\ problem exploration \\ and support\end{tabular} & \begin{tabular}[c]{@{}c@{}}Divide speaker roles \\ and initiative types\end{tabular} & \begin{tabular}[c]{@{}c@{}}External knowledge \\ (COMET, HEAL),\\  Graph Retrieval\end{tabular} & \begin{tabular}[c]{@{}c@{}}PPL, BLEU-4, \\ ROUGE-L\end{tabular} & \begin{tabular}[c]{@{}c@{}}Fluency, Identification, \\ Comforting, Suggestion\end{tabular} \\ \hline
Zhou et al. ~\cite{zhou2023facilitating} & 2023 & \begin{tabular}[c]{@{}c@{}}Absence of explicit \\ guidance for emotional \\ state improvement\end{tabular} & \begin{tabular}[c]{@{}c@{}}Evoking the seeker’s \\ emotion intensity\end{tabular} & \begin{tabular}[c]{@{}c@{}}MoE, External knowledge \\ (COMET, NRC-VAD), \\ RL\end{tabular} & \begin{tabular}[c]{@{}c@{}}PPL, BLEU-2,\\ DIST-2, cES, tES, \\ cDC, fDC\end{tabular} & \begin{tabular}[c]{@{}c@{}}Fluency, Informativeness, \\ Coherence, Supportiveness\end{tabular} \\ \hline
Zhao et al.~\cite{zhao2023transesc} & 2023 & \begin{tabular}[c]{@{}c@{}}Fail to fine-grained \\ dialogue turn transitions\end{tabular} & \begin{tabular}[c]{@{}c@{}}Capturing the transition \\ of strategy, emotion, \\ semantics\end{tabular} & \begin{tabular}[c]{@{}c@{}}External knowledge \\ (COMET, ATOMIC),\\  State Transition Graph, \\ Cross-attention\end{tabular} & \begin{tabular}[c]{@{}c@{}}ACC, PPL,\\  DIST-2, BLEU-4, \\ ROUGE-L\end{tabular} & \begin{tabular}[c]{@{}c@{}}Fluency, Identification, \\ Suggestion, Empathy\end{tabular} \\ \bottomrule
\end{tabular}%
}
\caption{ Overview of publications utilizing the ESConv ~\cite{liu2021towards}.}
\label{tab:app_esconv}
\end{table}

%% file: 4-experiment4-LLM.tex
\input{tab_llm}

\subsection{Post-LLM Approaches: Large Language Model–Based Dialogue Systems}~\label{sec:qual}

LLMs have demonstrated excellence in engaging in human-like interactions and following instructions to provide contextually relevant feedback~\cite{duan-etal-2024-botchat}. These abilities make LLMs suitable not only for general applications but also for specialized domains such as mental health~\cite{lawrence2024opportunities}, particularly in counseling systems~\cite{na-2024-cbt}. Their capability to manage complex, multi-turn dialogues allows for nuanced, empathetic, and adaptive interactions~\cite{bao2024large,duan-etal-2024-botchat}. 
Here, we investigated the applications and limitations of LLMs within dialogue systems that emulate counseling environments. To this end, we filtered papers from the entire collection that included specific author keywords, such as `large language models,' `GPT-3,' `ChatGPT,' `GPT-4,' and `GPT-3.5.' This filtering process yielded a total of 10 papers for examination as illustrated in Table~\ref{tab:app_llm}.

\noindent\textbf{\textit{Types and Attributes of Applied LLMs}}

In this section, we examined the specific LLMs utilized in each study and their distinct characteristics. Our findings revealed that the choice of LLM often varies based on its attributes and the intended application purpose.
Particularly, GPT-3~\cite{brown2020language} was the most prevalent LLM, applied in five of the ten papers~\cite{llanes2024developing,kharitonova2024incorporating,kaysar2023mental,qian2023harnessing,lee2022does}, while two papers~\cite{zhang-etal-2023-ask,qian2023harnessing} utilized GPT-3.5~\cite{roumeliotis2023chatgpt}.
GPT-3~\cite{brown2020language}, an autoregressive language model with 175 billion parameters, is trained on a comprehensive 570GB dataset to generate human-like text. This capability allows it to excel in tasks such as text generation, translation, summarization, and question-answering. GPT-3.5~\cite{roumeliotis2023chatgpt}, an upgraded version of GPT-3, is developed to advance language comprehension and generation abilities, delivering increased accuracy, efficiency, and adaptability across a range of natural language tasks. We posit that the selection of these models is likely due to their superior generative abilities and ease of use. 
In contrast, \citet{kharitonova2024incorporating} chose the open-source models Llama~\cite{touvron2023llama} and Llama2~\cite{touvron2023llama2}, which support local customization, to mitigate potential cost and privacy issues associated with the closed-source GPT series.
Meanwhile, LLMs fine-tuned for specific languages, such as Chinese, were exploited, including CHATGLM-6B~\cite{glm2024chatglm}, WenZhong~\cite{fengshenbang}, and PanGu~\cite{zeng2021pangu}.

\noindent\textbf{\textit{Techniques for Optimizing LLM}}

We studied the specialized techniques that distinguish LLMs from previous deep learning models. Our analysis identified zero-shot and few-shot learning as the frameworks that are predominantly utilized for LLMs. These approaches involved prompting methods that provide the model with either no examples or a few examples of questions and answers, enabling it to effectively address similar tasks. \citet{qian2023harnessing} supplied GPT-3 with random examples of empathetic responses and relevant background knowledge to support empathetic response generation. Likewise, both \citet{lee2022does} and \citet{firdaus2023multi} leveraged few-shot learning to produce empathetic responses, with the former focusing on aligning responses closely with the input query and the latter enhancing response empathy through emotion recognition.
Due to the generalizability of LLMs, they can attain high performance without further training, offering advantages over traditional deep learning models that demand vast amounts of training data~\cite{brown2020language}---a notable limitation in the mental health domain due to data scarcity. An alternative approach is Retrieval-Augmented Generation (RAG), a technique that integrates information retrieval with language generation. RAG initially retrieves pertinent information from a knowledge base and subsequently employs the LLMs to generate a response informed by this retrieved content~\cite{lewis2020retrieval}. \citet{kharitonova2024incorporating} applied this approach by retrieving knowledge items to respond to input queries in a psychological context. Comparably, \citet{chen2023emotional} implemented ChatGPT to produce multiple response candidates and used its reasoning capabilities to select the most suitable strategy for providing emotional support.

%% file: tab_llm.tex
\begin{table}[tbp]
\resizebox{\linewidth}{!}{%
\begin{tabular}{>{\centering\arraybackslash}p{1.6cm}|>{\centering\arraybackslash}p{0.8cm}|>{\centering\arraybackslash}p{2cm}|>{\centering\arraybackslash}p{3cm}|>{\centering\arraybackslash}p{2.5cm}|>{\centering\arraybackslash}p{2.5cm}|>{\centering\arraybackslash}p{3cm}|>{\centering\arraybackslash}p{2.2cm}|>{\centering\arraybackslash}p{2cm}}
\hline
\textbf{Reference} &
  \textbf{Year} &
  \textbf{Task} &
  \textbf{Psychological Background} &
  \textbf{Computation Approach} &
  \textbf{LLM} &
  \textbf{Dataset} &
  \textbf{Automatic Evaluation Metrics} &
  \textbf{Human Evaluation Metrics} \\  \hline
  
Lai et al.\cite{lai2023supporting} &
  2024 &
  Mental Health QA &
  - &
  Pretraining on corpus, Finetuning &
  WenZhong~\cite{fengshenbang} , PanGU~\cite{zeng2021pangu} &
  PsyQA~\cite{sun2021psyqa} &
  PPL, DIST-1, DIST-2, ROUGE-L & - \\
   
Llanes- Jurado et al.\cite{llanes2024developing} &
  2024 &
  Empathetic Response Generation &
  - &
  Providing context with GPT-3 followed by response generation. &
  GPT-3~\cite{brown2020language} &
  - &
  - &
  Naturalness, Realism, Valence, Arousal \\ 
  
Kharitonova et al.\cite{kharitonova2024incorporating} &
  2024 &
  Mental Health QA &
  - &
  Retrieval Augmented Generation(RAG) &
  GPT-3~\cite{brown2020language}, Llama~\cite{touvron2023llama}, Llama-2~\cite{touvron2023llama2}, &
  Synthetic QA from LLM &
  - &
  Coherence, Varacity, Evidence \\
  
Kaysar and Shiramatsu \cite{kaysar2023mental} &
  2023 &
  Providing suggestion for mental health problem &
  - &
  Natural Language Understanding (intent, emotion),  Finetuning &
  GPT-3~\cite{brown2020language} &
  Customized Conversational Datasets &
  BLEU, ROUGE &- \\
   
Firdaus et al. \cite{firdaus2023multi} &
  2023 &
  Empathetic Response Generation &
  - &
  Few-shot learning &
  DialogGPT~\cite{zhang-etal-2020-dialogpt} &
  DailyDialog~\cite{li2017dailydialog}, EmotionLines~\cite{hsu2018emotionlines}, EmoWOZ~\cite{feng2022emowoz} &
  BLEU, ROUGE-L & -\\
   
Lee et al. \cite{lee2022does} &
  2022 &
  Empathetic Response Generation &
  Expressing empathy requires emotional and cognitive insights~\cite{davis1980multidimensional} &
  Few-shot learning &
  GPT-3~\cite{brown2020language} &
  Empathetic Dialogues~\cite{rashkin2019towards} &
  DIST-2, NIDF, PPL, INTENTACC, EMOACC, Interpretation, Exploration, Emotion Reaction & -\\
   
Zhang et al. \cite{zhang-etal-2023-ask} &
  2023 &
  Emotional Support Conversation &
  - &
  Providing knowledge from GPT-3.5 as context &
  GPT-3.5~\cite{roumeliotis2023chatgpt} &
  ESConv~\cite{liu2021towards}, BlendedSkillTalks~\cite{roller-etal-2021-recipes} &
  BLEU-4, ROUGE-L, BERTScore, BARTScore & -\\
   
Chen et al. \cite{chen2023emotional} &
  2024 &
  Emotional Support Conversation &
  - &
  Finetuning, Voting &
  CHATGLM2-6B~\cite{glm2024chatglm} &
  ESConv~\cite{liu2021towards}(en) &
  BLEU-2, BLEU-4, DIST-1, DIST-2 &
  Empathy, Coherence, Helpfulness, Rapport \\
  
Qian et al.\cite{qian2023harnessing} &
  2023 &
  Empathetic Response Generation &
  - &
  Few-shot learning, Using knowledge-base for context &
  GPT-3~\cite{brown2020language}, GPT-3.5~\cite{roumeliotis2023chatgpt}, CHATGPT~\cite{roumeliotis2023chatgpt} &
  Empathetic Dialogue~\cite{rashkin2019towards} &
  DIST-1, DIST-2, P-BERTScore, R-BERTScore, F-BERTScore, BLEU-2, BLEU-4 &
  Fluency, Identification, Empathy, Coherence \\
  
Chen et al. \cite{chen2023soulchat} &
  2023 &
  Counseling Data Augmentation &
  - &
  Rewriting using ChatGPT, Finetuning &
  CHATGLM-6B~\cite{glm2024chatglm} &
  SoulChatCorpus SMILECHAT~\cite{qiu2024smile} &
  BLEU-1, BLEU-2, BLEU-3, BLEU-4, ROUGE-1, ROUGE-2, ROUGE-L &
  Naturalness,  Empathy, Helpfulness, Safety \\ \hline

\end{tabular}
}
\caption{Overview of research methodologies applied in studies utilizing LLMs.}
\label{tab:app_llm}
\end{table}

%% file: 5-Disucssion.tex
\section{Discussion}~\label{sec:discussion}
This section addresses the implications and recommendations for advancing mental health counseling systems. Additionally, it explores the study's limitations and outlines directions for future research.

\subsection{Implications \& Recommendations}

\subsubsection{The Lack of Training Dataset for Counseling System Development}
The difficulty of developing open-access datasets that represent the psychological counseling process remains a significant obstacle to advancing research in counseling system development. Our review indicates that existing studies have a constrained focus, often due to limited available datasets, when it comes to addressing the comprehensive counseling process. For example, among the ten most-cited papers, seven center exclusively on creating models intended to generate empathetic responses using the Empathetic Dialogue dataset~\cite{rashkin2019towards}, while many recent studies focus on providing emotional support through the ESConv dataset~\cite{liu2021towards}. 
Above all, the collection and annotation of counseling datasets is not only time-intensive and costly but also requires professional expertise~\cite{gibson2019multi}. Furthermore, due to confidentiality principles, counseling data cannot be shared externally without explicit client consent~\cite{younggren2008can,luepker2012record}. Since such data often contains personally identifiable information, anonymization through data preprocessing presents additional challenges~\cite{zuo2021data}.
In line with this, while ESConv is designed to simulate mental health counseling interaction, it encompasses only a narrow range of strategies, excluding techniques like Immediacy~\cite{kasper2008therapist} and Confrontation~\cite{moeseneder2019impact}---methods used by trained counselors to enhance therapeutic effectiveness---due to the lack of professional supervision and sufficient training resources. This constraint can lead to challenges in incorporating real-world scenarios into the model. 

To mitigate data scarcity, recent studies have increasingly aimed to apply LLMs to generate counseling datasets that encompass a wider variety of counseling strategies. Here, we provide a brief overview of four studies published after May 7, 2024, which were not included in our analysis. 
First, \citet{zheng2023augesc} fine-tuned GPT-J 6B~\cite{gpt-j} on 100 samples from the ESConv dataset and then leveraged the model to generate emotional support conversations by responding to dialogues from the Empathetic Dialogue dataset. This process yielded the AUGESC dataset, comprising 65,000 sessions across diverse topics. Experimental results and human evaluations confirmed that the dataset is non-toxic and closely emulates genuine emotional support conversations.
Second, \citet{zheng2024self} utilized ChatGPT~\cite{openai2023gpt4} to augment a conversation dataset by iteratively generating new data, using the ESConv dataset as a seed. This approach yielded the EXTES dataset, which includes 11,000 dialogues spanning 36 varied scenarios and 16 unique helping strategies. Human evaluations demonstrated that the quality of this dataset is comparable to that of the original ESConv dataset.
Third, \citet{zhang-etal-2024-cpsycoun} converted 3,134 high-quality psychological counseling reports into multi-turn consultation dialogues using a two-phase approach. In the first phase, counseling reports were transformed into clinical notes framed from the perspective of a psychological supervisor, offering guidance for the counselor. Based on these notes, simulated dialogues between a psychological counselor and client were then generated. 
Lastly, \citet{chen2023soulchat} introduced SoulChat, utilizing ChatGPT's text rewriting capabilities. By employing prompt-based transformations, they converted an initial dataset comprising 215,813 single-turn psychological counseling questions across 12 topics, along with 619,725 paired responses, into multi-turn conversations. This approach ultimately yielded a Chinese-language multi-turn empathetic conversation dataset containing 2,300,248 samples.

While LLM-driven data can support scalability in low-resource domains, it is important to note that LLMs still tend to exhibit unnatural patterns than human dialogue. Therefore, we suggest that future work incorporate more rigorous human-in-the-loop validation and linguistic analysis of LLM outputs to assess their clinical realism.

\subsubsection{The Need for Psychological Insights in Developing Counseling Systems} 

Psychological counseling process is grounded in extensive psychological knowledge and the specialized training of counselors~\cite{vacc2016professional,hill2020helping}.
However, our review indicated that while most studies prioritized accuracy improvements through computational techniques, they generally paid less attention to aligning their frameworks and underlying models with foundational psychological knowledge, which is crucial for practical application and usability. Moreover, several studies have drawn on empirical findings rather than psychological insights. For instance, \citet{tu2022misc} underscored the role of fine-grained emotion detection in emotional support based on empirical evidence alone. Likewise, \citet{cai-etal-2023-improving} prioritized commonsense knowledge over psychological expertise to enhance the system's comprehension of implicit social and emotional contexts.
These results also align with a recent review study that analyzed 69 studies employing LLMs in psychotherapy and found that only 32.8\% incorporated established psychotherapy theories into their methodology, indicating a gap between model development and theoretical grounding~\cite{na2025survey}. 
In contrast, \citet{liu2021towards} developed a dialogue dataset and trained a model that simulated the selection of context-sensitive conversational strategies, closely aligning actual counseling practices used by therapists. This approach, grounded in psychological knowledge, produced models that achieved high preference ratings in human evaluations.
Therefore, we propose that AI models for mental health improvement, particularly for interdisciplinary research and practical applications, be designed with a foundation in psychological knowledge. This integration is essential for accurately understanding individuals' psychological states and can greatly enhance the empathy quality exhibited by counseling systems~\cite{duan1996current}.

\subsubsection{Comprehensive Evaluation Metrics in Psychological Counseling Systems}

Evaluating the effectiveness of psychological counseling systems requires a careful examination of nuanced factors like empathy~\cite{clark2010empathy}, rapport~\cite{efstation1990measuring}, and perceived helpfulness~\cite{paulson1999clients}, all of which are critical for successful counseling~\cite{bachelor1991comparison}. However, as described in Table~\ref{tab:app_combined_metrics}, we found that many studies have concentrated on model performance, such as accuracy, text quality, and human-like fluency, while paying less attention to clinically relevant metrics for assessing the effectiveness of AI-driven counseling systems. 
While a growing number of studies have adopted human evaluations to overcome these limitations, such methods exhibit significant drawbacks. 
Human evaluations are time-consuming and lack consistency, as their results may vary depending on the evaluators involved, complicating comparisons across different counseling systems~\cite{hosking2024human,smith-etal-2022-human}. Additionally, they are prone to biases that may favor particular types of responses~\cite{moilanen2023review}.
Interestingly, recent research has explored the potential of using LLMs as evaluators in psychological counseling contexts~\cite{zheng2024judging}. For instance, \citet{zhang-etal-2024-cpsycoun} used GPT-4 to assess semantically complex factors, including comprehensiveness, professionalism, authenticity, and safety. Moreover, \citet{kang-etal-2024-large} examined biases in LLM-based dialogue systems toward specific emotional support strategies to achieve a balanced approach.
Nevertheless, an unmet need remains, suggesting that future directions for evaluation metrics should focus on establishing standardized and practical measures to facilitate the successful translation of AI systems into real-world psychological counseling settings.

However, an unmet need remains, suggesting that future directions for evaluation metrics should focus on establishing standardized and practical measures to facilitate the successful translation of AI systems into psychological counseling settings in the real world.

\subsubsection{Limitations of LLMs and Possibility of advancing LLMs in Psychological Counseling}
While LLMs have opened up vast possibilities in NLP, we identified major limitations when applying them to psychological counseling models. In this section, we reviewed existing approaches aimed at mitigating the limitations of LLMs and suggest how non-LLM-based research can inform future advancements.

\textbf{\textit{Hallucination}} emerged as a primary issue, defined as the phenomenon in which LLMs produce incorrect or irrelevant outputs in response to given inputs~\cite{ji2023towards,verspoor2024fighting}.
To reduce hallucination, additional information was supplied to support the generation of appropriate responses, such as incorporating emotionally similar contexts to align with input data~\cite{lee2022does,firdaus2023multi}. \citet{qian2023harnessing} trained the LLM using random contexts, enabling it to encounter and respond effectively to a range of scenarios.
In comparison, \citet{kharitonova2024incorporating} essentially assessed the risk of hallucination in LLMs by integrating a separate knowledge base. This approach enables the LLM to infer from carefully selected scenarios, thereby supporting the generation of safe responses. 
We further consider the integration of existing non-LLM approaches to handle hallucination by controlling over model outputs. For instance, reinforcement learning with reward functions has been employed to enhance empathetic responses~\cite{sharma2021towards,shin2020generating,li-etal-2020-empdg}. In addition, techniques such as incorporating strategy tokens~\cite{liu2021towards} and prior user feedback during strategy selection~\cite{peng2023fado,cheng2022improving} have been proposed to improve response consistency. These methods can offer promising solutions for reducing hallucinations in LLM-based dialogue systems.

\textbf{\textit{Limited mental health-related knowledge}} represents another critical limitation, as recent LLMs often demonstrate unreliability or inconsistency~\cite{agrawal2022large}, potentially due to inadequate understanding of mental health domains~\cite{yang2023towards}. Therefore, exploring approaches to strengthen the domain knowledge of LLMs is essential.
To address this limitation, in this study, \textit{Domain adaptation} through prompting or fine-tuning was applied to facilitate the effective use of LLMs as counseling models, primarily due to their limited exposure to relevant data during pretraining~\cite{wang-etal-2024-role,gururangan2020don}. 
In particular, prompting techniques are employed to initially identify the interlocutor's emotions~\cite{kaysar2023mental,firdaus2023multi,qian2023harnessing} or intentions~\cite{kaysar2023mental} to establish contextual understanding. \citet{lee2022does} also applied few-shot learning techniques, using examples that closely align with similar emotional and situational characteristics. This approach enables the model to condition responses based on the detected emotions or intentions, resulting in more appropriate replies. 
Additionally, \citet{lai2023supporting} trained their model on a comprehensive psychology corpus and fine-tuned it using PsyQA~\cite{sun2021psyqa} to enhance adaptation to the psychological domain. PsyQA is a Chinese dataset designed for generating long-form counseling responses in mental health support, consisting of 22,000 questions and 56,000 structured answers.
\citet{zhang-etal-2023-ask} underscored the knowledge limitations of smaller LLMs and positioned the LLM as a knowledge expert, given its extensive knowledge base acquired through diverse sources during training~\cite{zhou2024lima}. 
Existing techniques from non-LLM approaches could be adapted to enhance LLM-based systems. For instance, when retrieving external knowledge, constructing emotion-centered knowledge graphs can amplify their utility by injecting additional information about affective relationships~\cite{deng2023knowledge,peng2022ijcai}. Other approaches include integrating an emotion detection layer prior to generation, enabling the model to ground its outputs in more contextually relevant information~\cite{cheng2022improving,kim2021perspective,liu2021towards}.

\textbf{\textit{Limited context length}}~\cite{wang2024beyond} restricts their ability to retain personalized events and information across extended, ongoing counseling sessions. Effective psychological counseling generally involves multiple sessions, making it essential to monitor and retain details of the client's emotional state, experiences, and relevant events throughout the therapeutic journey to enable personalized counseling~\cite{choi2010evaluation}. Although personalization in chatbots has also demonstrated the potential to enhance therapy outcomes~\cite{vossen2024effect}, most current psychological dialogue systems are often trained on single-session datasets, limiting their capacity to provide personalized therapy. In line with this, \citet{zhong2024memorybank} introduced MemoryBank, a method designed to enhance the long-term memory capabilities of LLMs by enabling recall of prior interactions and adaptation to user personality traits.

\subsubsection{Ethical Considerations in the Development of Safe Systems}
For the safe use of AI counseling systems, it is pivotal to address ethical considerations. Risks include potential privacy infringements and leakage of personal information during both training and inference stages~\cite{molli2022effectiveness}. Furthermore, algorithmic biases and limitations in data may lead to culturally insensitive care or the dissemination of misinformation~\cite{rangaswamy2024ai}, or generation of psychologically harmful content~\cite{timmons2023call}.
To this end, model development should adhere to recognized guidelines, such as the American Psychological Association's Code of Ethics~\cite{american2002ethical} and AI risk management framework from NIST~\cite{ai2024artificial}. In constructing datasets, researchers must account for regulations such as the General Data Protection Regulation (GDPR) that cover commercial use, scientific data handling, informed consent, data deidentification, and adherence to a code of conduct~\cite{regulation2018general}. Thorough ethical consideration and researcher responsibility are vital to creating a safe and reliable counseling system.

\subsection{Study Limitations \& Future Directions}
\noindent\textbf{Study Limitations:} 
This study was intentionally designed to focus on a specific period and set of data sources in order to examine the technological transition in AI-based dialogue systems for mental health. The literature search was limited to studies published between 2020 and May 2024, a period that reflects the rise and broad adoption of LLMs while remaining centered on text-based dialogue systems. Consequently, very recent studies published after the search cutoff date may not be reflected in the current analysis. In addition, this study drew on three major citation databases which collectively provide broad and curated coverage of interdisciplinary research spanning computer science and human–computer interaction. While other venues and technically oriented publications may exist outside these databases, the selected sources were chosen to ensure reproducibility and relevance to mental health applications. Finally, although 146 studies were identified through the systematic search process, a subset of approximately 30 representative papers was selected for qualitative analysis. This focused approach allowed for an in-depth examination of key methodological trends before and after the integration of LLMs. 

\noindent\textbf{Future Directions:} For future systematic review research, scrutinizing more recent studies will help capture the latest trends in AI-driven psychological counseling systems. Expanding the investigation of interdisciplinary collaboration between the fields of computer science and mental health will better align technological advances with mental health needs. In addition, conducting a broader qualitative analysis covering all 146 identified papers, or a larger sample, could provide deeper insights into emerging trends and ethical considerations, improving our understanding of the future direction of AI in advancing mental health.

%% file: 6-Conclusion.tex
\section{Conclusions}
This study presents a comprehensive review of AI-driven dialogue systems for mental health by combining quantitative bibliometric analysis with a qualitative trend review of studies published between 2020 and May 2024. By examining 146 papers retrieved from Web of Science, Scopus, and the ACM Digital Library, we provide an overview of how research activity has developed during a period marked by the rapid proliferation of LLMs.

Our findings reveal a clear technological transition from task-specific deep learning approaches to LLM-based dialogue systems. Pre-LLM research primarily focused on modeling empathy through structured architectures, psychological knowledge integration, and dataset-driven optimization. Conversely, post-LLM research demonstrates the growing use of general-purpose language models for emotional support, offering enhanced linguistic flexibility and adaptability while introducing new challenges related to reliability, domain specificity, and safety.

Through this transition-oriented analysis, we identify several critical challenges and opportunities for advancing AI-based counseling systems. Key directions include the integration of psychological expertise into model design, improved access to high-quality and ethically grounded counseling datasets, and the development of robust evaluation frameworks that combine automatic and human-centered assessments. While LLMs show substantial promise in reducing development complexity and expanding system capabilities, addressing their limitations remains essential for real-world mental health applications.

Overall, this review contributes to both the machine learning and psychology communities by clarifying current research trajectories and providing a structured roadmap for future development. By situating recent advances within a broader technological context, this work aims to support the design of more effective, reliable, and clinically meaningful AI-driven dialogue systems for mental health support.

%% file: 7-ack.tex
\section*{Funding sources}
This research was supported by the MSIT (Ministry of Science, ICT), Korea, under the Global Research Support Program in the Digital Field program) (RS-2024-00425354) and the Global Scholars Invitation Program (RS-2024-00459638), supervised by the IITP (Institute for Information \& Communications Technology Planning \& Evaluation).

\section*{Declaration of competing interest}
The authors declare that there were no conflicts of interest with respect to the authorship or the publication of this article.

\section*{CRediT authorship contribution statement}
Daeun Lee: Writing – original draft, Data curation, Visualization, Software, Formal analysis, Donje Yoo: Writing – original draft, Data curation, Visualization, Formal analysis. Migyeong Yang: Writing - Visualization, review \& editing, Jihyun An: Writing – review \& editing, Validation, Christine B. Cha: Writing – review \& editing, Jinyoung Han: Writing – review \& editing, Methodology, Conceptualization, Supervision, Funding acquisition, Resources

\section*{Declaration of Generative AI and AI-assisted technologies in the writing process}
The authors declare that generative AI and AI-assisted technologies only used in the writing process to improve the readability and language of the manuscript.

\section*{Data availability}
The list of all 146 studies included in this review is provided in the Supplementary Material.